\documentclass[12pt,a4paper]{article}
\usepackage[]{lmodern,mathpazo}
\usepackage{amssymb,amsmath}
\usepackage{ifxetex,ifluatex}
\usepackage{fixltx2e} % provides \textsubscript
\ifnum 0\ifxetex 1\fi\ifluatex 1\fi=0 % if pdftex
  \usepackage[T1]{fontenc}
  \usepackage[utf8]{inputenc}
\else % if luatex or xelatex
  \ifxetex
    \usepackage{mathspec}
  \else
    \usepackage{fontspec}
  \fi
  \defaultfontfeatures{Mapping=tex-text,Scale=MatchLowercase}
  
\fi
\IfFileExists{upquote.sty}{\usepackage{upquote}}{}
\IfFileExists{microtype.sty}{%
\usepackage{microtype}
\UseMicrotypeSet[protrusion]{basicmath} % disable protrusion for tt fonts
}{}
\usepackage[marginparwidth=3cm,vmargin=1in,hmargin=1in,a4paper]{geometry}
\makeatletter
\@ifpackageloaded{hyperref}{}{%
\ifxetex
  \usepackage[setpagesize=false, % page size defined by xetex
              unicode=false, % unicode breaks when used with xetex
              xetex]{hyperref}
\else
  \usepackage[unicode=true]{hyperref}
\fi
}
\@ifpackageloaded{color}{
    \PassOptionsToPackage{usenames,dvipsnames}{color}
}{%
    \usepackage[usenames,dvipsnames]{color}
}
\makeatother
\hypersetup{breaklinks=true,
            bookmarks=true,
            pdfauthor={Carsten Allefelda*; Kai Görgena; John-Dylan Haynesa,b**},
            pdftitle={Valid population inference for  information-based imaging:  From the second-level t-test  to prevalence inference},
            colorlinks=true,
            citecolor=blue,
            urlcolor=blue,
            linkcolor=magenta,
            pdfborder={0 0 0}
            ,hidelinks,}
\usepackage{fancyvrb}
\VerbatimFootnotes % allows verbatim text in footnotes
\usepackage{graphicx}
\let\Oldincludegraphics\includegraphics
\renewcommand{\includegraphics}[1]{\hspace*{\fill}\makebox[0pt]{\Oldincludegraphics{#1}}\hspace*{\fill}}
\title{Valid population inference for\\
 information-based imaging:\\
 From the second-level \(t\)-test\\
 to prevalence inference}
\author{Carsten Allefeld\textsuperscript{a*} \and Kai Görgen\textsuperscript{a} \and John-Dylan Haynes\textsuperscript{a,b**}}
\date{}
\usepackage[font=small,format=plain,labelfont=bf]{caption}
\usepackage{titling}
\pretitle{\begin{center}\LARGE\bfseries}
\posttitle{\par\end{center}\vskip 0.5em}
\preauthor{\begin{center}
\lineskip 0.5em%
\begin{tabular}[t]{c}}
\postauthor{\end{tabular}\par\end{center}}
\predate{\begin{center}}
\postdate{\par\end{center}}
\providecommand{\mathup}[1]{\mathrm{#1}}                % if not already defined, e.g. for beamer as \mathsf
\newcommand{\ud}{\mathup{d}}                            % differential
\DeclareMathOperator{\Hnull}{H_0}                       % null hypothesis
\let\dist\sim                                           % distributed as
\newcommand{\Dist}[1]{\mathcal{#1}}                     % distribution
\ifx\paragraph\undefined\else
\let\oldparagraph\paragraph
\renewcommand{\paragraph}[1]{\oldparagraph{#1}\mbox{}}
\fi
\ifx\subparagraph\undefined\else
\let\oldsubparagraph\subparagraph
\renewcommand{\subparagraph}[1]{\oldsubparagraph{#1}\mbox{}}
\fi

\begin{document}
\maketitle

\begin{small}
a. Bernstein Center for Computational Neuroscience, Berlin Center of Advanced Neuroimaging, Department of Neurology, and Excellence Cluster NeuroCure, Charité – Universitätsmedizin Berlin, Germany\\
b. Berlin School of Mind and Brain and Department of Psychology, Humboldt-Universität zu Berlin, Germany\\[0.5em]
Address for all affiliations: Charité-Campus Mitte, Philippstr. 13, Haus 6, 10115 Berlin, Germany\\[0.5em]
* E-mail: carsten.allefeld@bccn-berlin.de\\
** E-mail: haynes@bccn-berlin.de\\[0.5em]
Corresponding author: Carsten Allefeld, Tel. +49 30 2093 6766
\end{small}

\bigskip
\bigskip
\bigskip
\framebox[\textwidth][c]{\parbox{0.95\textwidth}{\bf\centering\medskip 
preprint of\\[0.5em]
C. Allefeld, K. Görgen, J.-D. Haynes. Valid population inference for information-based imaging: From the second-level \emph{t}-test to prevalence inference. \emph{NeuroImage}, 141: 378–392, 2016. doi:10.1016/j.neuroimage.2016.07.040\\[0.5em]
this version includes minor fixes and a note added after publication\\
2016-8-10
\medskip}} \newpage

\subsubsection*{Abstract}\label{abstract}
\addcontentsline{toc}{subsubsection}{Abstract}

In multivariate pattern analysis of neuroimaging data, ‘second-level’
inference is often performed by entering classification accuracies into
a \(t\)-test vs chance level across subjects. We argue that while the
random-effects analysis implemented by the \(t\)-test does provide
population inference if applied to activation differences, it fails to
do so in the case of classification accuracy or other ‘information-like’
measures, because the true value of such measures can never be below
chance level. This constraint changes the meaning of the
population-level null hypothesis being tested, which becomes equivalent
to the global null hypothesis that there is no effect in any subject in
the population. Consequently, rejecting it only allows to infer that
there are some subjects in which there is an information effect, but not
that it generalizes, rendering it effectively equivalent to
fixed-effects analysis. This statement is supported by theoretical
arguments as well as simulations. We review possible alternative
approaches to population inference for information-based imaging,
converging on the idea that it should not target the mean, but the
prevalence of the effect in the population. One method to do so,
‘permutation-based information prevalence inference using the minimum
statistic’, is described in detail and applied to empirical data.

\subsubsection*{Keywords}\label{keywords}
\addcontentsline{toc}{subsubsection}{Keywords}

information-based imaging, multivariate pattern analysis, t-test,
population inference, effect prevalence

\newcommand{\A}{\mathup{A}}
\providecommand{\B}{\mathup{B}}
\renewcommand{\B}{\mathup{B}}
\newcommand{\Hn}{\mathup{H}_0}

\DeclareRobustCommand{\[}{\begin{equation}}
\DeclareRobustCommand{\]}{\end{equation}}

\section{Introduction}\label{introduction}

Since the seminal work of Haxby et al. (2001), an increasing number of
neuroimaging studies have employed multivariate methods to complement
the established mass-univariate approach (Friston et al., 1995) to the
analysis of functional magnetic resonance imaging (fMRI) data, a field
now known as multivariate pattern analysis (MVPA; Norman et al., 2006).
Most MVPA studies use classification (Pereira et al., 2009) to examine
activation patterns; the accuracy of a classifier in distinguishing
activation patterns associated with different experimental conditions
serves as a measure of multivariate effect strength. Since the target of
MVPA is not a generally increased or decreased level of activation but
the \emph{information content} of activation patterns (cf. Pereira and
Botvinick, 2011), it has also been characterized as information-based
imaging and distinguished from traditional activation-based imaging
(Kriegeskorte et al., 2006).

Many methodological aspects of MVPA have already been discussed in
detail: what kind of classifier to use (Cox and Savoy, 2003; Norman et
al., 2006), whether to adapt parametric multivariate statistics instead
of classifiers (Allefeld and Haynes, 2014; Nili et al., 2014), how to
understand searchlight-based accuracy maps (Etzel et al., 2013), or how
classifier weights can be made interpretable (Haufe et al., 2014;
Hoyos-Idrobo et al., 2015). By contrast, the topic of population
inference based on per-subject measures of information content, i.e.~the
question whether an information effect observed in a sample of subjects
generalizes to the population these subjects were recruited from, has
not yet received sufficient attention (but see Brodersen et al., 2013).

In univariate analysis of multi-subject fMRI studies, the standard way
to achieve population inference is to perform a ‘second-level’ null
hypothesis test (Holmes and Friston, 1998). For each subject, a
‘first-level’ contrast (activation difference) is computed, and this
contrast enters a second-level analysis, a \(t\)-test or an ANOVA.
Specifically for a simple one-sided \(t\)-test vs 0, reaching
statistical significance allows to infer that the experimental
manipulation is associated with an increase of activation on average in
the population of subjects. This is interpreted in such a way that the
effect is ‘common’ or ‘stereotypical’ in that population (Penny and
Holmes, 2007, p. 156).

With the adoption of information-based imaging, it has become accepted
practice to apply the same second-level inferential procedures to the
results of first-level multivariate analyses, in particular
classification accuracy (see e.g. Haxby et al., 2001; Spiridon and
Kanwisher, 2002; Haynes et al., 2007): A classifier is trained on part
of the data and is tested on another part, using each part for testing
once (cross-validation), and the classification performance is
quantified in the form of an accuracy, the fraction of correctly
classified test data points. Applied for example to two different
experimental conditions, if there was no multivariate difference in the
data between conditions, the classifier would operate at ‘chance level’,
i.e.~it would on average achieve a classification accuracy of
\(50\,\%\). At the second level, accuracies from different subjects are
then entered into a one-sided one-sample \(t\)-test vs \(50\,\%\), in
order to show that the ability to classify above chance and therefore
the presence of an information effect is typical in the population the
subjects were recruited from.

In this paper we argue that despite of the seemingly analogous
statistical procedure, a \(t\)-test vs chance level applied to
accuracies cannot provide evidence that the corresponding effect is
typical in the population. In contrast to other criticisms of this use
of the \(t\)-test (see below), in our view the problem is not so much
that the estimation distribution of cross-validated accuracies is not
normal or even symmetric, or that a normal distribution model is
generally inadequate for a quantity bounded to an interval
\([0\,\%, 100\,\%]\). Rather, the problem is that other than estimated
accuracies, the \emph{true single-subject accuracy can never be below
chance level} because it measures an amount of information.\footnote{Note
  that in this paper we only discuss the standard case of MVPA where the
  pair of experimental conditions is the same for training and test
  data. In ‘cross-decoding’ (cf. Haynes and Rees, 2005a), where it is
  tested whether a classifier trained on one pair of conditions is able
  to extract information corresponding to another pair of conditions,
  below-chance true accuracies may be possible. Cross-decoding targets
  not just the presence of information, but also the degree to which its
  neurophysiological representation is invariant with respect to another
  experimental manipulation.} We will show that this restriction changes
the meaning of the \(t\)-test: It now tests the global null hypothesis
(Nichols et al., 2005) that there is \emph{no information in any subject
in the population}. As a consequence, achieving a significant test
result allows us only to infer that \emph{there are people in which
there is an effect}, but not that the presence of information
generalizes to the population. The argument does not only hold for
classification accuracy, but also for other ‘information-like’ measures.

The \(t\)-test on accuracies has been criticized before (Stelzer et al.,
2013; Brodersen et al., 2013) on the grounds that its distributional
assumptions are not fulfilled for cross-validated classification
accuracies. Such a distributional error invalidates the calculation of
critical values for the \(t\)-statistic and can therefore lead to an
increased rate of false positives. This problem may be solved by better
distribution models (Brodersen et al., 2013) or the use of
non-parametric statistics (Stelzer et al., 2013). Our criticism goes
significantly beyond that: Not only is the \(t\)-test quantitatively
wrong, but it effectively tests a null hypothesis that is qualitatively
different from its use with univariate statistics, with the consequence
that rejection of this null hypothesis no longer supports population
inference.

Please note that our criticism pertains specifically to a second-level
\(t\)-test applied to per-subject classification accuracies or similar
measures. It does not apply to the classification \emph{of} subjects,
e.g.~into different patient groups in medical applications (Sabuncu and
Van Leemput, 2012; Sabuncu, 2014), or to the classification of
condition-specific patterns \emph{across} subjects (Mourao-Miranda et
al., 2005). Moreover, it only concerns quantities that measure the
information content of data, but not related quantities like classifier
weights (Wang et al., 2007; Gaonkar and Davatzikos, 2013; Gaonkar et
al., 2015, see below).

The organization of the paper is as follows: In Part~\ref{problem} we
detail how a second-level \(t\)-test achieves population inference for
univariate contrasts. We then explain that MVPA measures are
‘information-like’ and show, both theoretically and using simulations,
that for such measures the \(t\)-test effectively tests the global null
hypothesis that there is no effect in any subject.
Part~\ref{alternative} reviews possible alternatives to the \(t\)-test
on accuracies, converging on the idea that population inference for
information-based imaging should target the proportion of subjects in
the population with an effect. One way to implement such an ‘information
prevalence inference’ is described in detail in Part~\ref{method}, and
results of its application to real data are compared with those of the
\(t\)-test. We conclude with the discussion of a number of questions
surrounding the problem of population inference for information-based
imaging.

\section{\texorpdfstring{The problem with the \(t\)-test on
accuracies}{The problem with the t-test on accuracies}}\label{problem}

\subsection{Population inference in univariate fMRI
analysis}\label{population-inference-in-univariate-fmri-analysis}

To see why the \(t\)-test on accuracies cannot provide population
inference, we briefly recapitulate how standard univariate analysis does
achieve it. In a single subject, an activation difference or contrast
\(\Delta \beta\) is estimated based on the general linear model (GLM;
Friston et al., 1995). Because it is obtained from noisy data, the
estimate is itself noisy, \[ \label{conest}
\widehat{\Delta \beta} \dist \Dist N(\Delta \beta, \sigma_1^2),
\] where \(\sigma_1^2\) denotes the \emph{estimation variance} of the
contrast (cf.~Fig.~1a). If several subjects are included in a study, the
true activation difference \(\Delta \beta\) varies across subjects
(Fig.~1a): \[ \label{conpop}
\Delta \beta_k \dist \Dist N(\Delta \mu, \sigma_2^2)
\] where \(\Delta \mu\) is the average true activation difference in the
population of subjects and \(\sigma_2^2\) the \emph{population variance}
of the effect (Fig.~1b). The added subscript \(k\) indicates that we now
consider the subject as randomly sampled from the population. The
estimated contrast in several subjects therefore shows variation for two
reasons — they are noisy estimates (\(\sigma_1^2\)), and different
subjects respond differently (\(\sigma_2^2\)): \[ \label{conpopest}
\widehat{\Delta \beta}_k \dist \Dist N(\Delta \mu, \sigma_1^2 + \sigma_2^2).
\] The symbol \(\widehat{\Delta \beta}_k\) indicates that this contrast
is both estimated and sampled.

A one-sided \(t\)-test applied to the \(\widehat{\Delta \beta}_k\) from
a sample of subjects \(k = 1 \ldots N\) has the null hypothesis
\(\Delta \mu = 0\). If it can be rejected (\(\Delta \mu > 0\)), this
allows us to make a statement about the population of subjects because
\(\Delta \mu\) is a parameter of a \emph{population model}
(Eq.~\ref{conpop}). And this statement concerns a typical effect because
\(\Delta \mu\) is the mean, median, and mode of the assumed normal
distribution. This kind of test is also called \emph{random-effects
analysis} (RFX) because it treats subjects as as randomly sampled from a
population (Searle et al., 1992). It was introduced into fMRI by Holmes
and Friston (1998) to replace previous \emph{fixed-effects analyses}
(FFX), which did not account for population variation and therefore did
not provide population inference.\footnote{The specific RFX procedure to
  apply a second-level \(t\)-test or ANOVA to first-level contrast
  estimates is called the ‘summary statistic’ approach (Penny and
  Holmes, 2004) because it considers only the contrast estimates
  \(\widehat{\Delta \beta}_k\) which \emph{summarize} single-subject
  data. It is interesting to note that it is only this summary statistic
  approach that suggests other first-level summary statistics like
  classification accuracies could simply be plugged into a second-level
  \(t\)-test. For related attempts at a mixed-effects model for
  classification accuracies in a Bayesian setting see Olivetti et al.
  (2012) and Brodersen et al. (2012, 2013).}

\begin{figure}[htbp]
\centering
\includegraphics{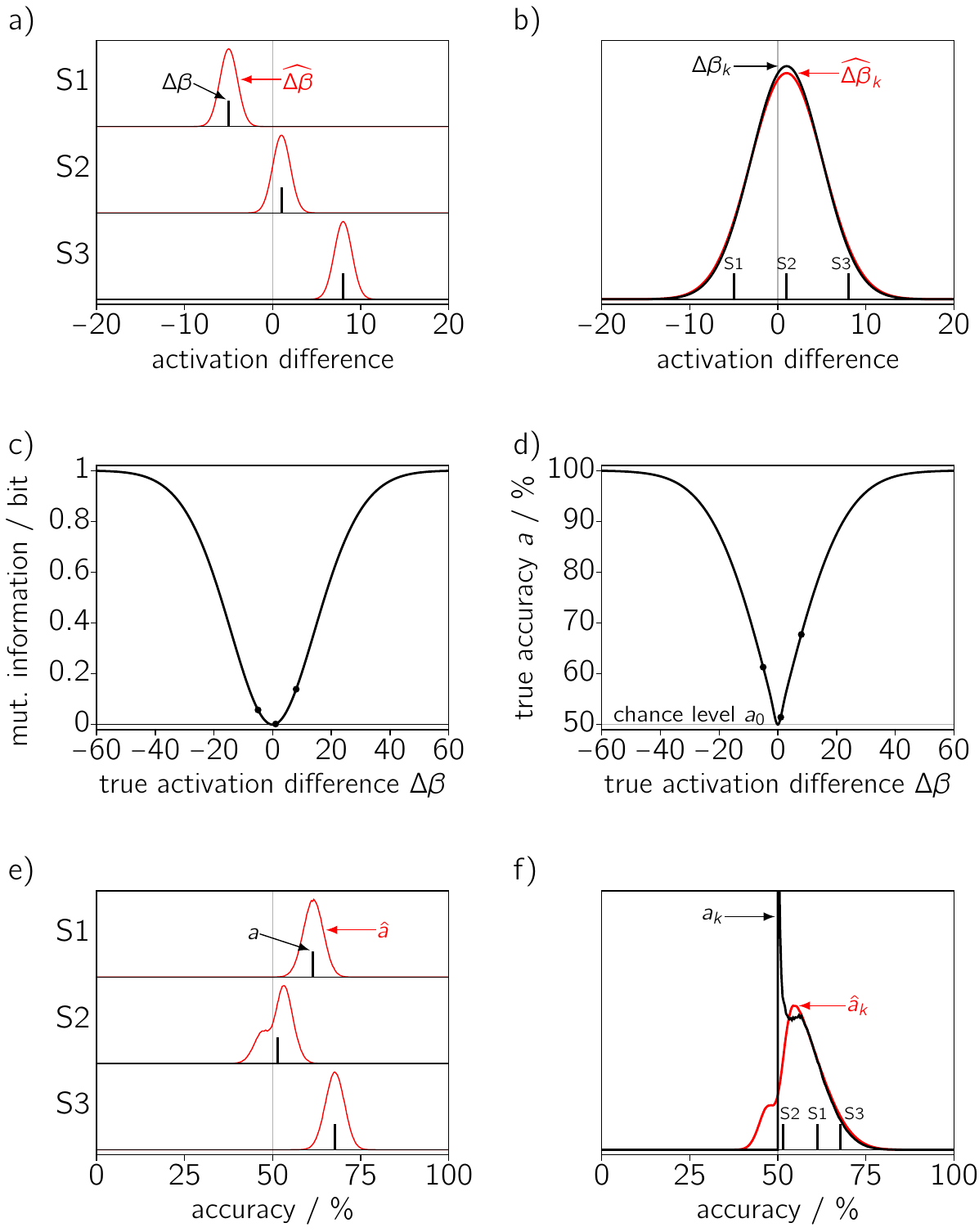}
\caption{Distributions of true (black) and estimated (red) values of
contrasts and classification accuracies. — a)~An activation difference
estimated from a limited amount of noisy data shows variation (red
curves; \(\sigma_1 = 1\)) around the true value (black bars). Moreover,
the true contrast is different in different subjects (Three panels S1,
S2, S3; \(\Delta\beta = -5,\,1,\,8\). Values for these subjects are also
indicated in the following panels by dots or bars.) — b)~True activation
differences (black bars) come from a distribution characterizing the
population of possible subjects (black curve; \(\Delta\mu = 1\),
\(\sigma_2 = 4\)). Estimated contrasts across subjects show the combined
effect of both sources of variation (population + estimation, red curve;
\({\scriptstyle \sqrt{\sigma^2_1 + \sigma^2_2}} = 4.12\)). — c)~The
amount of information single-trial data provide about the trial class
(condition) and vice versa, as a function of the true activation
difference. A negative contrast provides positive information. — d) True
accuracy \(a\) of classification of univariate single-trial data as a
function of the true activation difference. It is a symmetric function
of \(\Delta \beta\), which makes accuracy an information-like measure,
with a minimal value \(a_0 = 50\,\%\). — e)~Estimation variation of
accuracy (6-fold cross-validation) in the three subjects (red curves).
As apparent for subject 2, the distributions can deviate strongly from
normality. While estimated accuracies can be below chance (gray line),
true accuracies (black bars; \(a = 67.7,\,51.5,\,61.3\,\%\)) cannot. —
f)~The population variation (black curve) and combined variation
(population + estimation, red curve) of accuracy that result from the
population distribution of contrasts \(\Delta \beta_k\) (b, black line),
the functional relationship between \(\Delta\beta\) and \(a\) (d), and
the estimation distributions (e). The population distribution is
restricted to \(a \geq a_0\) and in this example shows a spike at
\(50\,\%\) and a weaker maximum at \(56\,\%\). — For further details,
see App.~\ref{simulation}.}
\end{figure}

\subsection{\texorpdfstring{The \(t\)-test on
accuracies}{The t-test on accuracies}}\label{the-t-test-on-accuracies}

Using a second-level \(t\)-test vs chance level with classification
accuracies implies that an analogous random-effects model applies: In
each subject (first level) we obtain an \emph{estimated accuracy} which
varies with an estimation variance \(\varsigma_1^2\), \[ \label{accest}
\hat a \dist \Dist N(a, \varsigma_1^2).
\] The underlying \emph{true accuracy} \(a\) (see App.~\ref{true})
varies across subjects (second level) with a population variance
\(\varsigma_2^2\), \[ \label{accpop}
a_k  \dist \Dist N(\bar a, \varsigma_2^2).
\] Therefore estimated accuracies vary across subjects with the combined
variance \(\varsigma_1^2 + \varsigma_2^2\), \[ \label{accpopest}
\hat a_k \dist \Dist N(\bar a, \varsigma_1^2 + \varsigma_2^2).
\] Here \(\bar a\) is the average true classification accuracy in the
population, and the null hypothesis is that this population average is
at chance level, \(\bar a = a_0\). Again, the symbol \(\hat a_k\)
indicates that this accuracy is both estimated and sampled. Though a
normal distribution cannot hold exactly since both \(\hat a\) and
\(a_k\) are limited to \([0\,\%, 100\,\%]\), we can argue that the
\(t\)-test is robust against violations of normality (Rasch and Guiard,
2004).

It appears that we have a viable random-effects model to justify the
application of a second-level \(t\)-test to accuracies. But there is a
problem with this simple transfer: In contrast to estimated accuracies
\(\hat a\), the true accuracy \(a\) can \emph{never be below} the chance
level \(a_0\).

\subsection{There is no negative
information}\label{there-is-no-negative-information}

To understand why \(a \geq a_0\) must hold, it helps to recognize that
MVPA aims at a generic kind of effect, namely, \emph{whether or not
information about experimental conditions is present in the experimental
data} (Pereira and Botvinick, 2011). Instead of making the common
distinction between univariate and multivariate fMRI analysis, we follow
Kriegeskorte et al. (2006; 2007) in distinguishing between
activation-based and information-based imaging. Activation-based
analysis is interested in whether there is a specific change in the
activation of a voxel (average BOLD signal) corresponding to an
experimental manipulation, normally an increase. Information-based
analysis determines whether there is any change at all, be it an
increase or decrease. It looks for brain areas where the difference of
conditions makes any kind of difference with respect to the fMRI signal,
i.e.~for information (Bateson, 1972).

Because information-based analysis disregards the sign of activation
differences, it has itself an unsigned outcome: There either is a
difference, then there is above-zero information, or there is no
difference, then there is zero information (‘chance level’). This holds
for information-theoretic measures in the strict sense (cf. Cover and
Thomas, 2012), in particular mutual information (Fig.~1c), but also for
the true value of \emph{information-like} measures including averaged
absolute \(t\) and Mahalanobis distance \(\Delta\) (used by Kriegeskorte
et al., 2006), Wilks’ \(\Lambda\) (used by Haynes and Rees, 2005b),
linear discriminant \(t\) (LD-\(t\), Nili et al., 2014), pattern
distinctness \(D\) (Allefeld and Haynes, 2014), or classification
accuracy \(a\).\footnote{We define as ‘information-like’ those measures
  that covary with mutual information. Note that this excludes other
  quantities that may be of interest in MVPA, in particular classifier
  weights (see Gaonkar and Davatzikos, 2013). Correspondingly, a
  classifier weight can be negative (cf. Todd et al., 2013), in the
  simplest case if information is coded in the corresponding voxel by a
  decrease in activation (but cf. Haufe et al., 2014).} The \emph{true}
single-subject accuracy is either above chance level, \(a > a_0\), if it
is possible to extract information, or it is at chance level,
\(a = a_0\), if not — but never below (Fig.~1d). \emph{Estimated}
accuracies can be below chance, but only due to imprecise estimation of
\(a\) by \(\hat a\), i.e.~below-chance accuracies are accounted for by
the first-level model of Eq.~\ref{accest}, not the second level of
Eq.~\ref{accpop}.

\subsection{\texorpdfstring{What does a \(t\)-test on accuracies
mean?}{What does a t-test on accuracies mean?}}\label{mean}

This restriction creates a problem for the second-level null hypothesis,
which \emph{qualitatively} changes its meaning. If \[ \label{meanchance}
\Hn: \bar a = a_0
\] is true, then \(a_k \dist \Dist N(a_0, \varsigma_2^2)\)
(Eq.~\ref{accpop}), which means that while half the people in the
population exhibit true above-chance classification, the other half is
assumed to systematically exhibit true below-chance classification,
contradicting our insight that \(a \geq a_0\). The null hypothesis of
the \(t\)-test can only be made compatible with this constraint by
additionally assuming that there is no population variation at all,
\(\Hn: \bar a = a_0 ~ \land ~ \varsigma^2_2 = 0\).\footnote{We would
  like to thank an anonymous reviewer for pointing out that this poses
  an additional distributional problem for the use of the \(t\)-test:
  The standard derivation of the null distribution of test statistics
  becomes invalid when the null hypothesis lies on the border of the
  parameter space (see Fahrmeir et al., 2013, Sec. 7.3.4).} And this
means that the the true accuracy is at chance level for everybody,
\[ \label{global}
\Hn: \forall_k ~ a_k = a_0,
\] — there is no information in \emph{any subject} in the population.
Such a null hypothesis, which is a logical conjunction of many simpler
null hypotheses, has been called ‘global null hypothesis’ by Nichols et
al. (2005).

Note that this conclusion does not depend on the assumption of normality
in Eq.~\ref{accpop} (which entered through the analogy to
Eq.~\ref{conpop}): For \emph{any} distribution of true accuracies, if
its mean is at chance but none of its realizations can be below chance,
it follows that there can be no above-chance realizations either.
Therefore the only possible form in which the null hypothesis formulated
in Eq.~\ref{meanchance} can hold is given by Eq.~\ref{global}.

The seemingly small constraint \(a \geq a_0\) has strong consequences
for inference. If the \(t\)-test allows us to reject the null
hypothesis, this provides evidence for the alternative, its logical
negation. Since the global null hypothesis (Eq.~\ref{global}) is a
universal statement, its negation is a statement of existence: \[
\neg \Hn: \exists_k ~ a_k > a_0.
\]

This means we have reason to believe that \emph{there are some} people
in the population whose fMRI data carry information about the
experimental condition — but we have no grounds to believe that we have
found an effect that is \emph{typical} in the population. The constraint
on \(a\) neutralizes the RFX modeling of between-subject variation,
making the \(t\)-test applied to accuracies effectively an FFX
analysis.\footnote{Todd et al. (2013) point out a related but different
  problem in applying a second-level test to a summary statistic that
  estimates an unsigned quantity: The traditional strategy of balancing
  or randomizing confounds across subjects (cf. Fisher, 1935) does no
  longer work, because after removing signs confounding effects of
  different direction cannot cancel each other out.}

Since the constraint that the true value cannot be below chance level
applies to all information-like measures, the problems for population
inference demonstrated here hold for information-based imaging in
general. However, in the interest of conciseness our discussion focuses
on cross-validated classification accuracy as the most commonly used
measure in MVPA.

At this point, the main argument of this paper is concluded: The
\(t\)-test on accuracies does not provide population inference, but
effectively implements fixed-effects analysis. The following section
further illustrates and practically demonstrates this fact using
simulated data. Readers which are already convinced by the theoretical
argument may skip forward to Part~\ref{alternative} which discusses
information prevalence inference as an alternative to the \(t\)-test.

\begin{figure}[htbp]
\centering
\includegraphics{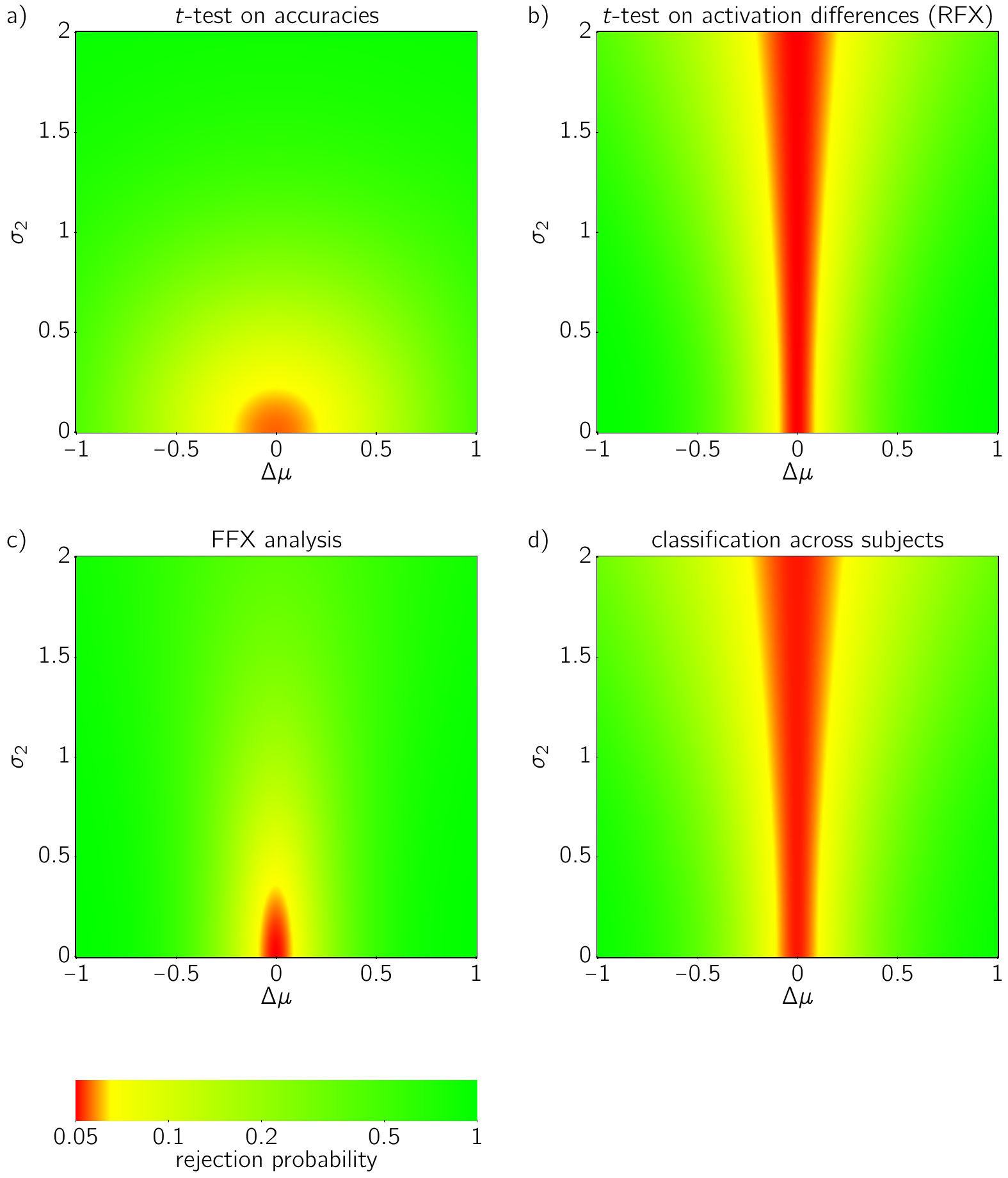}
\caption{Rejection probability as a function of simulation parameters
\(\Delta \mu\) (the population mean true contrast) and \(\sigma_2\) (the
population variance of true contrasts), for different null hypothesis
tests at significance level \(\alpha = 0.05\). — a)~Second-level
one-sided one-sample \(t\)-test applied to estimated classification
accuracies vs chance level \(a_0 = 50\,\%\). The smallest rejection
probability is reached if both \(\Delta \mu = 0\) and \(\sigma_2 = 0\).
— b)~Second-level two-sided one-sample \(t\)-test applied to estimated
activation differences vs 0 (univariate RFX analysis). The smallest
rejection probability is reached for \(\Delta \mu = 0\). —
c)~Fixed-effects analysis of activation differences. The smallest
rejection probability is reached if both \(\Delta \mu = 0\) and
\(\sigma_2 = 0\). — d)~Test based on classification across subjects. The
smallest rejection probability is reached for \(\Delta \mu = 0\).}
\end{figure}

\subsection{\texorpdfstring{What does a \(t\)-test on accuracies
do?}{What does a t-test on accuracies do?}}\label{do}

In the previous sections we gave a theoretical argument that the null
hypothesis of a second-level \(t\)-test changes its meaning under the
constraint that holds for information-like measures including
classification accuracy. But does this argument have practical relevance
— after all, we never see ‘true accuracies’ but only estimates, which
\emph{can} be below chance? Does our observation actually affect how a
\(t\)-test on accuracies \emph{behaves}?

To investigate this question, we simulate fMRI data according to the
first-level GLM (see Eq.~\ref{glm}) of a simplified experiment
containing two experimental conditions, for a sample of subjects. A
simulation is parametrized by the population mean activation difference
between conditions, \(\Delta \mu\), and the population variation
\(\sigma_2\) (Eq.~\ref{conpop}). The estimation variation
(Eq.~\ref{conest}) is kept at \(\sigma_1 = 1\). Accuracies for the
classification of single-trial data by a linear support vector machine
are estimated using run-wise cross-validation, and these accuracies are
entered into a one-sided one-sample \(t\)-test vs chance level
\(a_0 = 50\,\%\) across subjects at \(\alpha = 0.05\). This is repeated
many times to determine the \emph{probability to reject} the null
hypothesis. For full details of the simulations, see
App.~\ref{simulation} and \ref{implementation}.

\subsubsection{Classification of univariate data with a normal
population
model}\label{classification-of-univariate-data-with-a-normal-population-model}

For simplicity, we first examine the rejection probability of the
\(t\)-test on accuracies for the classification of univariate data;
Fig.~1 illustrates the distributions arising in this case. The resulting
rejection probability as a function of the simulation parameters
\(\Delta \mu\) and \(\sigma_2\) is shown in Fig.~2a.

The rejection probability function is a standard tool to check whether a
test is valid for a given null hypothesis and how powerful it is
(Lehmann and Romano, 2005). Here, we use the function in the opposite
direction: We \emph{define} the test’s \emph{effective null hypothesis}
as that set of parameter values where the rejection probability remains
at or below the specified significance level \(\alpha\).

For the \(t\)-test on accuracies (Fig.~2a) the result is that strictly
there are no such parameter values: the smallest rejection probability
is 0.055. This is most likely because the normality assumption of
Eq.~\ref{accpopest} holds only approximately; the slight increase of the
\(\alpha\)-error is however within the bounds used by Rasch and Guiard
(2004) when stating that the \(t\)-test is robust against violations of
normality. If we disregard this deviation, the effective null hypothesis
turns out to be \(\Delta \mu = 0 ~ \land ~ \sigma^2_2 = 0\): no
population variation, and no activation effect in any subject. This
means that the true activation difference is zero in all subjects,
\(\forall_k ~ \Delta \beta_k = 0\), and since the true accuracy
corresponding to a zero activation difference is at chance level,
\(a(\Delta \beta = 0) = a_0\), it is equivalent to
\(\Hn: \forall_k ~ a_k = a_0\) — no information in any subject in the
population. The simulation confirms the practical relevance of our
previous theoretical conclusion that the \(t\)-test on accuracies tests
the global null.

For comparison, Fig.~2b shows the rejection probability function of a
second-level two-sided \(t\)-test on first-level activation differences
(univariate RFX analysis). In accordance with the design of the test,
its rejection probability remains at 0.05 if and only if
\(\Delta \mu = 0\), and increases monotonically with \(|\Delta \mu|\).
Though \(\sigma^2_2\) is not part of the specification of the null
hypothesis, we observe that for stronger population variation it becomes
harder to reach a significant effect. This is in stark contrast to the
\(t\)-test on accuracies (Fig.~2a), where increasing \(\sigma^2_2\)
makes it easier to reject the null hypothesis. The more
\emph{inconsistent} activation differences (or by extension: patterns;
see below) are in the population, the more likely it is that the
\(t\)-test on accuracies will indicate the presence of an information
effect that is supposedly ‘typical’ in the population!\footnote{Davis et
  al. (2014) make a similar observation, but without noting its
  consequences for population inference.}

As we noted above, univariate fixed-effects analysis does not provide
population-level inference simply because its null hypothesis (that the
\emph{sample} mean is 0) does not reference a population distribution.
We can however apply an FFX test to our simulated univariate RFX data
and thereby determine the null hypothesis it effectively implements in
this context. The result shown in Fig.~2c demonstrates that the
effective null hypothesis of FFX analysis is
\(\Delta \mu = 0 ~ \land ~ \sigma^2_2 = 0\). Though their rejection
probability functions are different in detail, a \(t\)-test on
accuracies and a fixed-effects analysis of activation differences
operate qualitatively identical: they both detect deviations from a zero
population mean contrast \(\Delta\mu\) as well as from a zero population
variance of constrasts \(\sigma^2_2\). This agreement supports our
earlier statement that even though a \(t\)-test on accuracies formally
acknowledges population variation, it does not provide population
inference any more than FFX analysis does.

To complete the picture, Fig.~2d shows the rejection probability
function of a test based on \emph{classification across subjects} (cf.
Mourao-Miranda et al., 2005). The single-subject univariate contrast
estimates in each of the two conditions form the data set that is used
to train and test a classifier in leave-one-subject-out
cross-validation, and the distribution of accuracies is determined for
different simulation parameters. For a given simulated data set the null
hypothesis (true accuracy = chance level) is rejected if the accuracy
reaches or exceeds the critical value of \(67.6\,\%\), which is the
approximate 95th percentile of the distribution of accuracies under the
null hypothesis. The result demonstrates that this test behaves in a way
that is very similar to the second-level \(t\)-test applied to the same
first-level activation differences (Fig.~2b). Both provide population
inference because they both implement the null hypothesis
\(\Delta \mu = 0\) for arbitrary population variance \(\sigma^2_2\).

The difference is that while the second-level \(t\)-test is limited to
univariate contrasts, the test based on classification across subjects
can just as well be applied to multivariate data, where the null
hypothesis becomes \(\Delta \vec \mu = \vec 0\). If a significant effect
is found, this provides evidence that there is a \emph{pattern
difference} \(\Delta \vec \mu\) that is typical in the population.

Since this implies that the presence of information is typical in the
population, it appears that the problem of population inference for
information-based imaging could simply be solved by applying classifiers
or other MVPA methods always across subjects. Unfortunately, spatial
normalization algorithms do not achieve precise voxel-level anatomical
alignment (Thirion et al., 2006), and moreover we cannot assume that a
one-to-one correspondence of informative patterns between subjects
always exists (Haynes and Rees, 2006; Kriegeskorte and Bandettini, 2007;
Haxby, 2012), so that classification across subjects is often bound to
fail for a trivial reason.\footnote{A solution might be provided by
  ‘hyperalignment’ (Haxby et al., 2011) which attempts to establish a
  fine-grained \emph{functional} correspondence between different
  subjects’ brains. However, as Todd et al. (2013) point out, aligning
  patterns effectively discards sign (direction) information, too,
  unless the hyperalignment parameters are determined from separate
  data.}

\begin{figure}[htbp]
\centering
\includegraphics{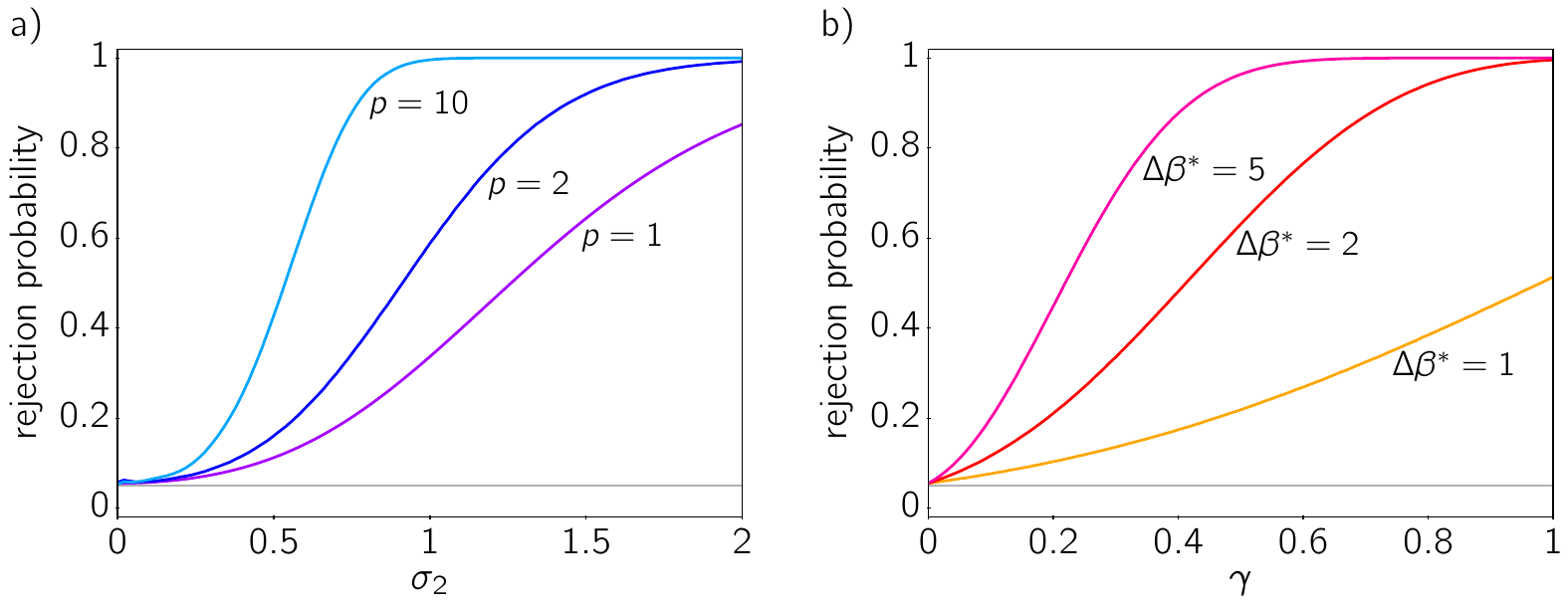}
\caption{Rejection probability of a second-level one-sided one-sample
\(t\)-test applied to estimated classification accuracies vs chance
level, as a function of simulation parameters. The significance level
\(\alpha = 0.05\) is shown as a gray horizontal line. — a)~Multivariate
normally distributed contrasts in \(p\) voxels, with variation
uncorrelated between voxels and a population mean activation difference
of \(\Delta\mu = 0\) everywhere. The rejection probability increases
with the population variance \(\sigma^2_2\), and the increase is
stronger for higher dimensionality \(p\). The line for univariate data
(\(p = 1\)) corresponds to a central vertical section through the
rejection probability function of Fig.~2a. — b)~Population proportion
model: Fixed true contrast \(\Delta\beta^*\) in a proportion \(\gamma\)
of the population, and 0 in the rest. The rejection probability always
reaches \(\alpha\) for \(\gamma = 0\).}
\end{figure}

\subsubsection{Classification of multivariate data with a normal
population
model}\label{classification-of-multivariate-data-with-a-normal-population-model}

An important observation in the last section was that the larger the
population variance \(\sigma^2_2\), i.e.~the more inconsistent
activation differences are across subjects, the easier it becomes for
the \(t\)-test on accuracies to achieve significance. To make sure that
this effect is not limited to univariate data, Fig.~3a shows the
rejection probability function for different numbers of dimensions,
\(p = 1, 2\), and \(10\). Data are generated as before, but for \(p\)
different voxels in parallel, with both first- and second-level
variation uncorrelated between voxels. Accordingly, the classifier is
trained and tested in a \(p\)-dimensional space. To facilitate
visualization, we kept the population mean activation difference at zero
in all voxels, \(\Delta \mu = 0\). The result demonstrates that the
effect of population variance on the rejection probability stays
qualitatively the same, but is even stronger in higher dimensions. We
can therefore assume that the effective null hypothesis of the
\(t\)-test on accuracies generalizes from the univariate
\(\Delta \mu = 0 ~ \land ~ \sigma^2_2 = 0\) to the multivariate
\(\Delta \vec \mu = \vec 0 ~ \land ~ \Sigma_2 = 0\), where \(\Sigma_2\)
is the multivariate population variance (variance–covariance matrix).
That implies \(\forall_k ~ \Delta \vec \beta_k = \vec 0\) — there is no
informative pattern in any subject in the population — which is again
equivalent to \(\Hn: \forall_k ~ a_k = a_0\), the global null.

\subsubsection{Classification of univariate data with a population
proportion
model}\label{classification-of-univariate-data-with-a-population-proportion-model}

Our simulation-based finding that the effective null hypothesis of a
\(t\)-test applied to first-level accuracies is
\(\Hn: \forall_k ~ a_k = a_0\) (the global null) was so far obtained
using the standard normal distribution population model for activation
differences (Eq.~\ref{conpop}), which may not be correct (cf. Rosenblatt
et al., 2014). With respect to true accuracies, this model for true
contrasts has a peculiar consequence: As soon as there is any population
variation, i.e. \(\sigma^2_2 > 0\), the probability that the true
contrast in a given subject \(k\) is exactly \(\Delta\beta_k = 0\)
becomes zero, which implies that almost always
\(\forall_k ~ \Delta\beta_k \neq 0\). Since for non-zero true contrast
the true accuracy is above chance, \(a(\Delta\beta \neq 0) > a_0\)
(cf.~Fig.~1d), this further implies that almost always
\(\forall_k ~ a_k > a_0\). Therefore the assumption of a normal
population distribution of activation differences allows only two
possibilities: Either there is no information in the data of \emph{any}
subject (the effective null hypothesis), or there is information in the
data of \emph{every} subject. There is nothing in between.

To see how the \(t\)-test on accuracies reacts to a situation between
these extremes, we use an alternative population model for activation
differences (replacing Eq.~\ref{conpop}; modified from Rosenblatt et
al., 2014): Assume that the true contrast has a fixed value
\(\Delta\beta^*\) in a certain proportion \(\gamma \in [0, 1]\) of the
population and the fixed value 0 in the rest. If now a subject \(k\) is
randomly selected from the population, this means that
\[ \label{simpleprop}
\Delta\beta_k = \left \{
\begin{array}{ll}
0               & \text{with probability } 1 - \gamma \\
\Delta\beta^*   & \text{with probability } \gamma
\end{array}
\right ..
\] Fig.~3b shows the behavior of the \(t\)-test in a simulation using
this model. The result is that for different values of
\(\Delta\beta^*\), the rejection probability reaches the significance
level \(\alpha\) always at \(\gamma = 0\). This effective null
hypothesis is again equivalent to the global null,
\(\Hn: \forall_k ~ a_k = a_0\). Moreover, the simulation demonstrates
that the \(t\)-test on accuracies may with high probability declare an
information effect to be ‘typical’ even though it is only present in a
small minority of subjects in the population!

\section{An alternative: Information prevalence
inference}\label{alternative}

In the previous part we established that the second-level \(t\)-test
applied to accuracies is not able to provide population inference. We
now discuss alternative approaches, leading us to the idea that
population inference for information-based imaging should target the
\emph{proportion} of people in the population in which there is an
information effect.

Within the MVPA literature, there are three alternative proposals.
First, Kriegeskorte and Bandettini (2007) recommend to apply the methods
for ‘combining brains’ collected by Lazar et al. (2002). However, except
for the summary-statistic approach of Holmes and Friston (1998), all of
these methods\footnote{Fisher’s (1925) combined probability test,
  Tippett’s (1931) minimum \(p\)-value, the conjunction test of Worsley
  and Friston (2000), Stouffer et al.’s (1949) combined \(z\)-value,
  Mudholkar and George’s (1979) logit method, and fixed-effects
  analysis.} are meta-analytic procedures which explicitly test the
global null. Second, Stelzer et al. (2013) propose to generate
single-subject permutation statistics, and then to construct a
second-level permutation distribution by randomly selecting first-level
permutations in each subject. Because each permutation realizing the
second-level null hypothesis is a combination of permutations realizing
the first-level null hypotheses in every subject, this again tests the
global null hypothesis of no information in any subject. And third,
Brodersen et al. (2013) follow Olivetti et al. (2012) and Brodersen et
al. (2012) in describing a mixed-effects analysis for MVPA, introducing
explicit estimation and population models for classification accuracies.
This approach offers several improvements over the \(t\)-test on
accuracies: Their model can account for different estimation variances
(\(\varsigma^2_1\)) in different subjects, and it uses more realistic
distributional assumptions. However, the authors do not consider the
fact that true accuracies are limited to the range \([a_0, 1]\) (for
binary classification: \([50\,\%, 100\,\%]\)). Unfortunately, this
renders their approach only an improved version of the \(t\)-test on
accuracies.

As detailed in Sec.~\ref{mean}, the flaw of the \(t\)-test on accuracies
lies in the fact that the distributional assumption of the underlying
second-level model (Eq.~\ref{accpop}) is incompatible with the
restriction \(a \geq a_0\), unless \(\varsigma_2 = 0\). This could
conceivably be fixed by using another population model which adheres to
that restriction by design. However, the simulated population
distribution of true accuracies for univariate classification in Fig.~1f
does not look like it could be appropriately captured by a parametric
model. And even if that were the case, the distribution would likely
have a different shape for classification in higher dimensions or for
multi-class classification. It might additionally depend on the specific
classification algorithm, and it would certainly be different again for
other information-like measures.

Moreover, inference with respect to the population mean is generally
inadequate for information-like measures no matter how well the actual
population distribution can be modeled. The reason is that the true mean
is above chance as soon as there is a true above-chance effect in a
small fraction of the population (cf.~Fig.~3b; because there can be no
true below-chance effect). This problem can be resolved by inference
that does not target the mean effect, but the proportion of subjects in
the population in which there is an information effect, i.e.~the
\emph{prevalence} of information.

Such an approach is followed by Rouder et al. (2007) in the context of
signal detection theory with their ‘mass-at-chance model’. They assume a
probit-normal population distribution of true detection accuracies,
which however is truncated at \(50\,\%\), such that all subjects that
are nominally below chance are instead located at chance. Inference with
respect to the prevalence of an effect has also been advocated by
Rosenblatt et al. (2014) for mass-univariate analysis. The authors argue
that imperfect alignment of single-subject activation maps leads to
areas where only a subset of subjects have a non-zero activation, and
propose to replace the standard normal population model
(Eq.~\ref{conpop}) by a mixture model. We used a simplified version of
this in the population proportion simulation (Eq.~\ref{simpleprop} and
Fig.~3b). Stephan et al. (2009) propose a Bayesian form of prevalence
inference as RFX analysis for dynamic causal models (DCM), extending
first-level model selection to a posterior distribution over the space
of different model frequencies. In a specific application this discrete
distribution may describe the distinction between a zero effect and a
generic non-zero effect.\footnote{Stephan et al.’s Bayesian RFX analysis
  for DCM has been adapted for GLM model selection by Soch et al.
  (2016), and may be adapted for MGLM (cf. Allefeld and Haynes, 2014)
  model selection to support MVPA.} Friston et al. (1999a) build upon
their previous idea of a conjunction test (Price and Friston, 1997;
Worsley and Friston, 2000) and introduce the minimum-statistic approach.
Their analysis proceeds in two steps: first, the minimum statistic is
used to derive a \(p\)-value for the null hypothesis that there is no
effect in any subject in the population, the global null. In a second
step, a correction is applied that allows to test the null hypothesis
that the proportion of subjects in which there is an effect, \(\gamma\),
is at or below a threshold \(\gamma_0\). The rejection of this null
hypothesis therefore allows to infer that \(\gamma > \gamma_0\).

The approaches of Rouder et al. (2007), Rosenblatt et al. (2014),
Stephan et al. (2009), and Friston et al. (1999a) are all candidates to
be adapted for information-based imaging, to provide population
inference with respect to the prevalence of an information effect. In
the following we demonstrate this in detail for the method of Friston
and colleagues.

\section{Permutation-based information prevalence inference using the
minimum statistic}\label{method}

In this part we recapitulate the minimum-statistic approach to
prevalence inference developed by Friston et al. (1999a), adapt it to be
based on permutation statistics, and detail the resulting algorithm.
Applied to information-like measures this method allows us to achieve
\emph{information prevalence inference}, i.e.~inference with respect to
the proportion of subjects in the population that exhibit an information
effect. We demonstrate the method using an example data set.

The advantage of Friston et al.’s approach is that it can be implemented
based on known permutation methods at the single-subject level (Golland
and Fischl, 2003; Etzel and Braver, 2013; Schreiber and Krekelberg,
2013; Stelzer et al., 2013; Allefeld and Haynes, 2014; see also Ernst,
2004). The method is discussed with respect to classification accuracy,
but the test logic applies equally to other information-like measures.

A note on notation: We previously used variables without index (e.g.
\(a\)) when talking about the ‘first level’ of a given single subject,
and symbols with index (e.g. \(a_k\)) when considering a subject as
randomly selected from the population, or referring to all members of
the population (\(\forall_k\)). This notation is still followed, but we
now extend it such that an index associated with an explicit range,
\(k = 1 \ldots N\), refers to the specific subjects included in a given
sample.

\subsection{The minimum statistic and the global
null}\label{the-minimum-statistic-and-the-global-null}

In a single subject, an estimated classification accuracy \(\hat a\) has
an associated \(p\)-value \(p(\hat a)\), which is the probability to
observe an accuracy that large or larger given that the true accuracy is
at chance level (\(a = a_0\)).

For a sample of \(N\) subjects with estimated classification accuracies
\(\hat a_k\), \(k = 1 \ldots N\), we choose the \emph{minimum statistic}
(smallest observed accuracy) as the second-level test statistic,
\[ \label{minimum}
m = \min_{k = 1}^N \hat a_k.
\] As a second-level null hypothesis we first consider the \emph{global
null hypothesis}, \[ \label{globalagain}
\Hn: \forall_k ~ a_k = a_0,
\] that there is no effect in any subject in the population. In order to
test this null hypothesis, we need the \(p\)-value of \(m\), \(p_N(m)\),
with respect to the global null.

To say that the minimum of estimated accuracies is at or larger than a
given value \(m\) is the same as saying that all of the estimated
accuracies \(\hat a_k\), \(k = 1 \ldots N\), are at or larger than
\(m\). Since subjects are independently drawn from the population, the
probability to observe a minimum of \(m\) or larger under the global
null is the product of probabilities to observe an estimated accuracy of
\(m\) or larger in each subject in the sample: \[ \label{puncglo}
p_N(m) = \prod_{k = 1}^N p(m) = p(m) ^ N,
\] where \(p(m)\) is the single-subject \(p\)-value for \(\hat a = m\).

If \(p_N(m) \leq \alpha\), then we can reject \(\Hn\) and infer that
there are some subjects in the population in which there is an
above-chance effect. Since this is a statement of existence, this does
not provide evidence that the effect is typical in the population.

\subsection{The prevalence null}\label{the-prevalence-null}

We now consider a population model which contains the global null as a
special case: The information effect targeted by the classification
procedure has a \emph{prevalence} \(\gamma\), i.e.~a proportion
\(\gamma \in [0, 1]\) of subjects in the population have an above-chance
effect, the others no effect. If a subject \(k\) is selected at random
from this population, then \[
\begin{array}{ll}
a_k = a_0   & \text{with probability } 1 - \gamma, \\
a_k > a_0   & \text{with probability } \gamma.
\end{array}
\] This is similar to the population proportion model for activation
differences used above (Eq.~\ref{simpleprop}), but in contrast no
assumption is made about the size and distribution of above-chance
effects.

An estimated accuracy \(\hat a\) in a single subject can be larger than
or equal to \(m\) either purely by chance (\(p(m)\), with probability
\(1 - \gamma\)) or because there is actually an effect in that subject
(\(p(m | a > a_0)\), with probability \(\gamma\)). The probability to
observe a sample minimum of \(m\) or larger if the prevalence is
\(\gamma\) is therefore \[
\begin{split}
p_N(m | \gamma) &= \prod_{k = 1}^N [ (1 - \gamma) ~ p(m) + \gamma ~ p(m | a > a_0) ] \\
&= [ (1 - \gamma) ~ p(m) + \gamma ~ p(m | a > a_0) ] ^ N.
\end{split}
\] Here \(p(m | a > a_0)\) is the probability to observe an estimated
accuracy of \(m\) or larger in a single subject given a true accuracy of
\(a\), where we only know that \(a > a_0\). Because the size of the
above-chance effect \(a > a_0\) is not specified, we cannot know this
probability precisely; but because it is a probability, we know that it
is smaller than or equal to one, \(p(m | a > a_0) \leq 1\), and
therefore \[ \label{ineq}
p_N(m | \gamma) \leq [ (1 - \gamma) ~ p(m) + \gamma  ] ^ N.
\] The reason we chose the minimum statistic as the second-level test
statistic is that it enables us to formulate this inequality for the
prevalence model, i.e.~to determine a \(p\)-value without specifying the
size and distribution of above-chance effects.

The prevalence model allows us to formulate the \emph{prevalence null
hypothesis}, \[
\Hn: \gamma \leq \gamma_0,
\] that the prevalence is smaller than or equal to a threshold
prevalence \(\gamma_0\). The global null (Eq.~\ref{globalagain}) is a
special case of the prevalence null where \(\gamma_0 = 0\).

The prevalence null hypothesis is a complex null hypothesis, i.e.~it can
be realized by different values of the parameter \(\gamma\). In such a
case, the \(p\)-value associated with a test statistic is the
probability to observe the given value or larger, \emph{maximized} over
all situations consistent with \(\Hn\). Therefore \[
p_N(m | \gamma \leq \gamma_0) = [ (1 - \gamma_0) ~ p(m) + \gamma_0  ] ^ N.
\] In a permutation approach, \(p_N(m)\) can be more precisely
determined than \(p(m)\) (see step~3 under Algorithm). We therefore
express the prevalence null \(p\)-value
\(p_N(m | \gamma \leq \gamma_0)\) in terms of the global null
\(p\)-value \(p_N(m)\), using the relation \(p_N(m) = p(m)^N\)
(Eq.~\ref{puncglo}): \[ \label{puncprev}
p_N(m | \gamma \leq \gamma_0) = [ (1 - \gamma_0) ~ \sqrt[N]{p_N(m)} + \gamma_0  ] ^ N.
\] If \(p_N(m | \gamma \leq \gamma_0) \leq \alpha\), then we can reject
\(\Hn\) and infer that the prevalence \(\gamma\) is significantly larger
than \(\gamma_0\), i.e.~more than a proportion \(\gamma_0\) of the
population have an effect.

As an alternative to fixing a threshold prevalence \(\gamma_0\) in
advance and then testing the corresponding prevalence null, we can
compute the largest \(\gamma_0\) such that the corresponding null
hypothesis can still be rejected at the given significance level
\(\alpha\): \[
\gamma_0 = \frac{\sqrt[N]{\alpha} - \sqrt[N]{p_N(m)}}{1 - \sqrt[N]{p_N(m)}}.
\] Note that this is not an estimator for the true prevalence
\(\gamma\), but \(]\gamma_0, 1]\) is a one-sided
\((1 - \alpha)\)-confidence interval for it.

This confidence interval will often be too wide because of the
inequality used above (Eq.~\ref{ineq}). Moreover, even for the strongest
possible effect (\(p_N(m) = 0\)) it holds
\(\gamma_0 = \sqrt[N]{\alpha}\), meaning that the strength of the
population inference is limited by the number of subjects \(N\) and the
chosen significance level \(\alpha\).

\subsection{Information maps}\label{information-maps}

So far we have considered a single second-level test based on
classification accuracies \(\hat a_k\) from \(N\) different subjects.
But in a common variant of MVPA, searchlight analysis (Kriegeskorte et
al., 2006), we have \emph{maps} of classification accuracies and perform
a test at each voxel in these maps, i.e.~we have to adjust for multiple
comparisons.

To do so, we need to specify a \emph{spatially extended} version of the
prevalence null. Again following Friston et al. (1999a), our spatially
extended null hypothesis is:\\
— there is an effect with prevalence \(\gamma \leq \gamma_0\) in a small
area,\\
— and no effect everywhere else.\\
The justification for this is that in experiments investigating the
localization of information we normally expect this information to be
restricted to specialized brain areas.

Under this null hypothesis, a sample minimum of \(m\) or larger can
occur either purely by chance (no true effect), with a probability that
is increased because we examine many voxels at once (\(p_N^*(m)\)); or
if that is not the case (\(1 - p_N^*(m)\)) it can occur because there
actually is an effect in a sub-threshold proportion of the population,
with a probability that is not increased because the effect is only
present in a small area (\(p_N(m | \gamma \leq \gamma_0)\)). Here
\(p_N^*(m)\) is the \(p\)-value for the spatially extended global null,
corrected for multiple comparisons using a standard method (see step~4
under Algorithm). Taken together, the probability to observe a sample
minimum of \(m\) or larger at a given voxel, corrected for multiple
comparisons according to the spatially extended prevalence null
hypothesis is \[ \label{pcorprev}
p_N^*(m | \gamma \leq \gamma_0) = p_N^*(m) + [1 - p_N^*(m)] ~ p_N(m | \gamma \leq \gamma_0).
\] For given threshold \(\gamma_0\), the spatially extended prevalence
null can be rejected at a particular voxel if
\(p_N^*(m | \gamma \leq \gamma_0) \leq \alpha\). Equivalently, we can
define a significance level that is corrected for multiple comparisons,
\[ \label{alphacor}
\alpha^* = \frac{\alpha - p_N^*(m)}{1 - p_N^*(m)},
\] and reject the spatially extended prevalence null if the uncorrected
\(p\)-value is at or below the corrected level,
\(p_N(m | \gamma \leq \gamma_0) \leq \alpha^*\). Note that because \(m\)
is voxel-specific, \(\alpha^*\) is, too.

Again, we can alternatively compute the largest \(\gamma_0\) such that
the corresponding spatially extended prevalence null can still be
rejected: \[ \label{confidence}
\gamma_0^* = \frac{\sqrt[N]{\alpha^*} - \sqrt[N]{p_N(m)}}{1 - \sqrt[N]{p_N(m)}},
\] which results in a map of lower confidence bounds of the prevalence
of the effect, an \emph{information prevalence map}.

\subsection{Algorithm}\label{algorithm}

We now explain in detail how the computations derived above can be
implemented based on first-level permutation statistics.

Step 1: For each subject, classification accuracies \(\hat a_v\) are
computed for each voxel \(v\). Additionally, classification accuracies
are computed for data where the class labels have been permuted,
\(\hat a_{vi}\) with \(i = 1 \ldots P_1\), where \(P_1\) is the number
of available first-level permutations. \(i = 1\) denotes the neutral
permutation, i.e. \(\hat a_{v1} = \hat a_v\).

Step 2: The minimum classification accuracy \(m_v\) across subjects
(Eq.~\ref{minimum}) is computed at each voxel \(v\). Additionally, the
minimum accuracy is computed for each second-level permutation,
\(m_{vj}\) with \(j = 1 \ldots P_2\). A second-level permutation is a
combination of first-level permutations, and therefore there are
\(P_1^N\) possible second-level permutations. If there are too many
combined permutations, a subset can be selected randomly (Monte Carlo
estimation), but it has to be made sure that \(j = 1\) denotes the
combination of first-level neutral permutations (\(i = 1\) in all
subjects). This procedure of combined permutations is identical to the
one employed by Stelzer et al. (2013).

Step 3: The uncorrected \(p\)-value for the global null hypothesis is
determined at each voxel (Eq.~\ref{puncglo}) as \[
p_N(m_v) = \frac1{P_2} \sum_{j = 1}^{P_2} [ m_v \leq m_{vj} ]
\] where the so-called Iverson bracket \([\cdot]\) has the value 1 for a
true condition and the value 0 for a false condition. That is,
\(p_N(m_v)\) is the fraction of combined-permutation values of the
minimum statistic larger than or equal to the actual value. Because
\(m_{v1} = m_v\), the smallest possible \(p\)-value is \(\frac1{P_2}\).

Step 4: To correct \(p_N(m_v)\) for multiple comparisons, the maximum
statistic across voxels is computed for each combined permutation (see
Nichols and Holmes, 2001), \[
M_j = \max_v m_{vj},
\] and then \[
p^*_N(m_v) = \frac1{P_2} \sum_{j = 1}^{P_2} [ m_v \leq M_j ]
\] is determined. \(p^*_N(m_v)\) is the \(p\)-value for the spatially
extended global null hypothesis.

Step 5a: To determine where the spatially extended prevalence null
hypothesis for a given threshold \(\gamma_0\) can be rejected, at each
voxel Eq.~\ref{puncprev} is used to compute
\(p_N(m_v | \gamma \leq \gamma_0)\) from \(p_N(m_v)\),
Eq.~\ref{pcorprev} to compute \(p_N^*(m_v | \gamma \leq \gamma_0)\) from
\(p^*_N(m_v)\) and \(p_N(m_v | \gamma \leq \gamma_0)\), and it is
checked whether \(p_N^*(m_v | \gamma \leq \gamma_0) \leq \alpha\).

Step 5b: Alternatively, to determine for each voxel the largest
threshold \(\gamma_0^*\) at which the spatially extended prevalence null
hypothesis can be rejected, Eq.~\ref{alphacor} is used to compute
\(\alpha^*_v\) from \(p^*_N(m_v)\). Then \(p_N(m_v) \leq \alpha^*_v\) is
checked to see whether the spatially extended global null hypothesis can
be rejected. If that is not the case, the largest threshold
\(\gamma_0^*\) is not defined for that voxel (the prevalence null cannot
be rejected, even at \(\gamma_0^* = 0\)). For all voxels where the
spatially extended global null hypothesis can be rejected,
Eq.~\ref{confidence} is used to compute \(\gamma_{0v}^*\) from
\(\alpha^*_v\) and \(p_N(m_v)\). Note that the maximally possible
\(\gamma_{0v}^*\) determined this way is limited not just by the chosen
significance level \(\alpha\) and the number of subjects \(N\), but also
by the number of second-level permutations \(P_2\); it is
\[ \label{gamma0max}
\gamma_{0\mathrm{max}}^*
= \frac{\sqrt[N]{\alpha^*_\mathrm{max}} - \sqrt[N]{1 / P_2}}{1 - \sqrt[N]{1 / P_2}}
\quad \text{with} \quad
\alpha^*_\mathrm{max} = \frac{\alpha - 1 / P_2}{1 - 1 / P_2}.
\] For large \(P_2\), \(\alpha^*_\mathrm{max}\) is approximately equal
to \(\alpha\).

A problem for this method may arise from the fact that both the minimum
statistic underlying prevalence inference and the maximum statistic used
to correct for multiple comparisons do not produce new values (unlike
e.g.~the mean, which in general differs from all the values it is
calculated from). Since the number of possible classification accuracies
is limited because of a limited number of data points, this may lead to
a large number of permutations where the statistic attains the same
value (tied permutation values), which inflates the \(p\)-values
computed in steps 2 and 3 above. This problem can be solved by using
spatially smoothed accuracy maps as inputs (which is also advisable to
reduce residual anatomical misalignment between subjects), or by using a
continuously-valued information-like measure like pattern distinctness
(Allefeld and Haynes, 2014) instead.

\subsection{Application}\label{application}

In order to illustrate our permutation-based implementation of
information prevalence inference, we re-use the data of Cichy et al.
(2011). Twelve different visual stimuli belonging to four different
categories were presented either to the left or the right of fixation
(24 experimental conditions) to \(N = 12\) subjects. There were four
different trials per condition in each of five different runs. fMRI data
were recorded from a field of view covering the ventral visual cortex at
an isotropic resolution of 2 mm. Data were preprocessed and normalized
to the MNI template. A linear SVM with parameter \(C = 1\) was trained
on GLM parameter estimates from four of the runs, and tested on the
fifth run, in a leave-one-run-out cross-validation scheme.
Classification was pairwise (\(24 \cdot 23 \, / \, 2 = 276\) pairs) and
accuracies were averaged across pairs combining different factor levels,
so that the chance-level accuracy was \(a_0 = 50\,\%\). For permutation
statistics, class labels were exchanged in each of the five runs
separately, which lead to \(P_1 = 2^{5 - 1} = 16\) unique first-level
permutations. The analysis was performed using a searchlight of radius 4
voxels (comprising 257 voxels). The resulting accuracy maps were
smoothed with a Gaussian kernel of \(6\,\mathrm{mm}\) FWHM. For more
details, see Cichy et al. (2011) and Allefeld and Haynes (2014). For
information prevalence inference, we randomly selected \(P_2 = 10^7\)
out of \(P_1^N = 2.8\cdot10^{14}\) possible combined permutations at the
second level. All the following results are corrected for multiple
comparisons, but for simplicity we omit the superscript ’\(^*\)’.

\begin{figure}[htbp]
\centering
\includegraphics{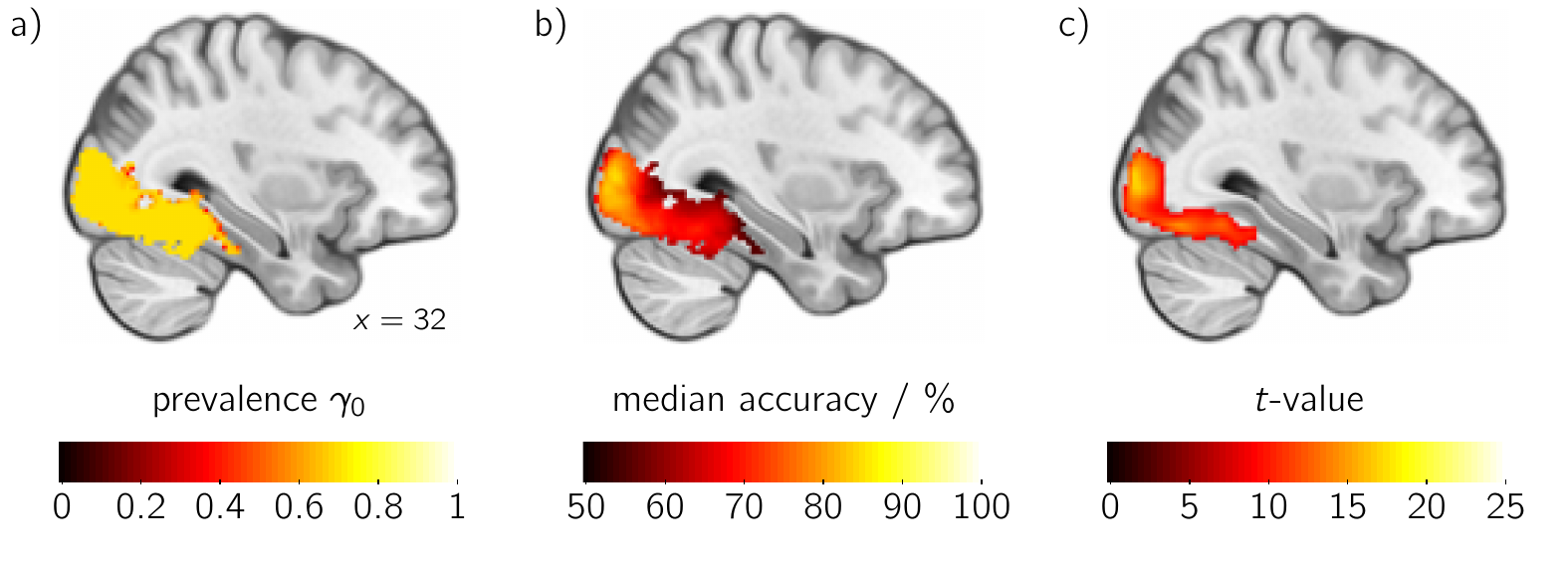}
\caption{Second-level results for the classification of object category
of a visual stimulus (see Cichy et al., 2011) shown in a sagittal slice
through right lateral occipital cortex and fusiform gyrus. Statistics
are corrected for multiple comparisons. — a)~Information prevalence
inference. Highlighted areas are those where the global null hypothesis
(prevalence \(\gamma = 0\)) can be rejected at a level of
\(\alpha = 0.05\). Colors visualize a lower bound \(\gamma_0\) on the
prevalence of category information (confidence level 0.95). — b)~For
those areas where the prevalence null hypothesis
\(\gamma \leq \gamma_0\) can be rejected at \(\gamma_0 = 0.5\),
i.e.~where it can be inferred that the majority of subjects in the
population have an effect, colors visualize the median classification
accuracy across subjects. — c)~\(t\)-test on accuracies vs chance level
\(a_0 = 50\,\%\). For those areas where the null hypothesis
\(\bar a = a_0\) can be rejected at a level of \(\alpha = 0.05\), colors
visualize the underlying \(t\)-value.}
\end{figure}

The results for the classification of stimulus category are shown in
Fig.~4. The spatially extended global null hypothesis of no information
in any subject in the population can be rejected at a level of
\(\alpha = 0.05\) in about \(27\,\%\) of in-mask voxels. For those
voxels, the largest lower bound \(\gamma_{0}\) at which the spatially
extended prevalence null hypothesis can be rejected is shown in Fig.~4a.
In about half of those voxels, \(\gamma_{0}\) reaches the maximally
possible value (for the given sample size \(N\), significance level
\(\alpha\), and number of second-level permutations \(P_2\)) of
\(\gamma_{0\mathrm{max}} = 0.701\) (Eq.~\ref{gamma0max}).

For those voxels where the largest lower bound \(\gamma_{0}\) is larger
than or equal to 0.5, i.e.~where we can infer that in the
\emph{majority} of subjects in the population the data contain
information about the stimulus category, the median estimated accuracy
is shown in Fig.~4b. We chose the median across all subjects in the
sample as a descriptive statistic to accompany our prevalence results,
because it can be considered as a cautious and robust estimator of the
typical above-chance classification accuracy.\footnote{For
  \(\gamma > 0.5\) we can consider subjects where the true accuracy is
  at chance (no information) as ‘outliers’. The median is a robust
  estimator with a breakdown point of 0.5, i.e.~it can handle samples
  where up to half of the values are outliers (Huber and Ronchetti,
  2009). It is cautious in the sense that it will under- rather than
  overestimate the the true median above-chance accuracy, because all
  ‘outliers’ are on the side of small values.}

For comparison, the result of a second-level \(t\)-test on accuracies vs
chance level is shown in Fig.~4c. Although the effective null hypothesis
of this test is identical to the global null hypothesis \(\gamma = 0\)
explicitly tested in Fig.~4a, the number of voxels at which it can be
rejected is much smaller (about \(14\,\%\) of in-mask voxels at
\(\alpha = 0.05\), FWE-corrected), indicating that in this case the
\(t\)-test is less sensitive. The picture is the same for the comparison
with the test of the prevalence null hypothesis \(\gamma \leq \gamma_0\)
at \(\gamma_0 = 0.5\) shown in Fig.~4b. Information prevalence inference
therefore allows us to draw conclusions that are stronger than those
provided by the \(t\)-test on accuracies, concerning both interpretation
— population inference — and, for this data set, statistical power.
However, the result of Fig.~4b also calls into question whether the
assumption that an effect is constrained to a ‘small area’ (adopted from
Friston et al., 1999a) is generally adequate.

\section{Discussion}\label{discussion}

In this paper we have shown that the \(t\)-test on accuracies commonly
used in MVPA studies is not able to provide population inference because
the true single-subject accuracy \(a\) can never be below chance level.
This constraint makes the effective null hypothesis of the test the
global null hypothesis that there is no effect in any subject in the
population, which means that in rejecting that null hypothesis we can
only infer that there are some subjects in which there is an effect.
This is in stark contrast to the standard interpretation of a
significant result of a second-level \(t\)-test. We supported our
statement both by theoretical arguments, in particular detailing that
classification accuracy in MVPA is an information-like measure, as well
as by simulations of the relevant distributions to investigate the
practical behavior of the \(t\)-test applied to accuracy data. Finally,
we reviewed possible alternative inference methods and described one
approach that can be implemented based on known first-level permutation
statistics in combination with the minimum statistic as the second-level
test statistic. In the following we discuss a number of possible
counter-arguments and questions.

\emph{Does this mean that the \(t\)-test fails, is not robust enough?} —
Not really. In all the instances we examined, the \(t\)-test does what
it is supposed to do: Check whether the population mean is increased.
The point is that under the constraint \(a \geq a_0\), rejecting
\(H_0: \bar a = a_0\) no longer tells us that the effect is typical in
the population; an increased accuracy in a small fraction of the
population is sufficient to increase the population mean (cf.~Fig.~3b).
Average information content is not typical information content.
Concerning robustness, for the case we simulated (Fig.~2a) the
\(\alpha\)-error was only slightly increased, consistent with the notion
that the \(t\)-test is indeed robust.

\emph{Is it wrong to test the global null hypothesis?} — No. The
decision which null hypothesis to test rests with the researchers
performing a study. The problem lies with the interpretation of a
significant result, which for second-level analysis — at least tacitly —
is to infer that the effect is typical in the population. Moreover, the
step from sample to population inference (FFX to RFX) as the gold
standard has been taken already a long time ago for univariate analyses
(Holmes and Friston, 1998) and more recently for DCM (Stephan et al.,
2009); it would therefore be natural to expect that it also becomes the
standard for information-based imaging. In particular, any claim that
MVPA is more sensitive than univariate analysis (Norman et al., 2006) is
meaningless if MVPA is not held up to the same inferential standards.

\emph{We observe below-chance accuracies all the time.} — There are two
aspects to this. For one, our statement that accuracies cannot be below
chance refers to \emph{true} accuracies, not to estimated accuracies.
Second, it is sometimes the case that estimated accuracies strongly
suggest that the true accuracy is below-chance, too.\footnote{With
  respect to fMRI–MVPA, this phenomenon is informally discussed (see
  e.g.~J. Etzel’s blog,
  \url{http://mvpa.blogspot.de/2013/04/below-chance-classification-accuracy.html}),
  but appears to not yet have given rise to a peer-reviewed publication.}
This is most likely due to the circumstance that a crucial assumption of
cross-validation is not met, namely that the different parts of the data
(here: fMRI recording sessions) come from the same distribution (Efron
and Tibshirani, 1994). If there are systematic changes across data
parts, for example because of a confound either in the data or in the
experimental design (Görgen et al., 2014), this can induce a negative
bias, including the possibility that classifier performance lies
systematically below chance. (For another tentative explanation, see
Kowalczyk, 2007.) This does of course not mean that now there is
‘negative information’, but only that cross-validated accuracy is not a
meaningful information measure under such circumstances. This
possibility does not invalidate the argument made in this paper, but
points to another problem that needs a separate remedy.

\emph{A classifier could be designed to systematically give the wrong
answer, leading to a true accuracy below chance.} — Yes, but in that
case the accuracy of this classifier would no longer be a measure of the
information content of the data. For example in the simple case where
the output of a working classifier is falsified by always returning the
opposite classification result (A instead of B and B instead of A),
information content would no longer be quantified by \(a - a_0\), but by
\(- (a - a_0)\). The argument in this paper does not refer to
classifiers in general, but to the use of classification accuracy and
other measures to quantify the information content of data.

\emph{What does it mean for an effect to be ‘typical’ in the
population?} — This question essentially asks about the scientific
content of statistical inference at the population level. Surprisingly,
the topic is almost never discussed in statistical scientific papers and
textbooks, including those aimed at psychologists and other cognitive
scientists. Our use of the term ‘typical’ was inspired by Penny and
Holmes (2007), who state that in population inference ‘one is interested
in what is common to the subjects’ or in a ‘stereotypical effect in the
population’ (p.~156). In this paper, we use the term mainly in a
negative way: An effect that is only present in a small fraction of the
population can hardly be considered typical (or ‘common’, or
‘stereotypical’). In standard univariate analysis concerning the mean of
a normal distribution population model, a positive use of the term can
be motivated by the fact that the mean is also the mode of the
distribution (the most frequent realization) as well as its median (the
value that sits ‘in the middle’ of the distribution). But there is
another aspect: If in this context we can reject the null hypothesis
\(\Hnull: \Delta \mu = 0\) in a one-sided \(t\)-test, the inference
\(\Delta \mu > 0\) also means that more than half of the population —
the \emph{majority} — have a positive effect (and less than half a
negative effect). This statement can be extended to the non-normal case
if there is a test that can show that the population \emph{median} is
above zero. If we take this observation as a guideline, the natural
choice for the threshold in prevalence inference is \(\gamma_0 = 0.5\)
(see also Friston et al., 1999b). If we can reject this null hypothesis,
we can again infer that there is an effect in the majority of subjects
in the population, which also implies that the median true effect
strength is above-chance (motivating Fig.~4b).\footnote{Additionally,
  this choice is consistent with the use of the ‘exceedance probability’
  in Bayesian RFX (Stephan et al., 2009), which is the posterior
  probability that a given model is more frequent in the population than
  all other models, if only two models are considered.} We therefore
propose to call an effect ‘typical’ if it is present in the majority of
subjects in the population.

\emph{But don’t we want to show that the effect generalizes in the sense
that it is present in every subject in the population?} — Maybe ideally,
but in practice this is impossible. First, it is not known whether any
of the effects that are of interest in functional neuroimaging do
actually exhibit such an extreme degree of generalization, in particular
with respect to a specific anatomical location. Second, statistical
inference is unable to provide support for such a statement on
principle. As pointed out above, a univariate one-sided \(t\)-test for a
normal distribution only provides evidence that a majority of subjects
in the population have an effect, not that it generalizes to everyone.
And in population prevalence inference, at best we could test the
prevalence null hypothesis for \(\gamma_0 = 0.99\) or similar, at the
price of very low sensitivity.

Concluding we would like to point out that we do not consider the method
of ‘permutation-based information prevalence inference using the minimum
statistic’ put forward in the last part to be the definitive solution to
the problem of population inference in information-based imaging. The
main aim of this paper was to demonstrate conclusively that the
\(t\)-test on accuracies does not provide population inference, and that
prevalence inference is an alternative. This particular method was
presented in order to show that population inference is possible, even
based on established methodology only (minimum statistic, first-level
permutations). However, it does have several shortcomings: The use of
the minimum statistic limits the highest \(\gamma_0\) for which the
prevalence null hypothesis can be rejected depending on the number of
subjects, and the permutation-based implementation imposes an even lower
limit depending on the number of second-level permutations that can be
performed. Moreover, this method does not provide a way to estimate
(instead of just bound) the true population prevalence \(\gamma\). We do
however believe that methods focusing on population prevalence are the
most promising approach, not just for information- but also for
activation-based imaging, because they explicitly provide information
about the \emph{degree} to which an effect generalizes. And while we
hope that this paper will motivate further methodological work on
population inference for information-based imaging, our method does
provide a way to improve upon the commonly used \(t\)-test on accuracies
that is available now.\footnote{An implementation for Matlab can be
  obtained from the corresponding author or at
  \url{http://github.com/allefeld/prevalence-permutation/releases}, and
  is also included in The Decoding Toolbox (TDT; Hebart et al., 2015)
  from version 3.8.}

\section*{Acknowledgments}\label{acknowledgments}
\addcontentsline{toc}{section}{Acknowledgments}

Kai Görgen was supported by the German Research Foundation (DFG grants
GRK1589/1 and FK:JA945/3-1).

The authors would like to thank Tom Nichols, Jakob Heinzle, Jörn
Diedrichsen, Will Penny, María Herrojo Ruiz, Joram Soch, Martin Hebart,
Jo Etzel, Yaroslav Halchenko, and Thomas Christophel for discussions,
comments, and hints.

\section*{Appendix}\label{appendix}
\addcontentsline{toc}{section}{Appendix}

\appendix
\renewcommand{\thesubsection}{\Alph{subsection}}

\subsection{What is a true accuracy?}\label{true}

The first-level model (Eq.~\ref{accest}) for the RFX analysis of
accuracies assumes that there is a ‘true accuracy’ \(a\) that underlies
the random estimated accuracies \(\hat a\) that we actually observe. But
what does this true value stand for?

In the case of a true activation difference \(\Delta \beta\), the answer
is simple: The first-level GLM \[ \label{glm}
y_t = \beta_\A \, x_{\A t} + \beta_\B \, x_{\B t} + e_t
\] where \(x_{\A t}\) and \(x_{\B t}\) are regressor functions and
\(e_t\) is noise, provides a statistical description of the (neural and
hemodynamic) process by which we assume the observed fMRI data are
generated (a generative model). This model includes parameters
\(\beta_\A\) and \(\beta_\B\) which govern the generative process, and
which characterize the respective subject. From a given data set
\(y_t\), we can estimate these parameters, \(\hat \beta_\A\) and
\(\hat \beta_\B\), but because the data are noisy, these estimates will
not coincide with the underlying true values. However, if we could
repeat the experiment an infinite number of times \emph{with the same
subject}, the mean of estimates across these repetitions would recover
the true values, because the estimators are unbiased (provided the
parameters are estimable). This mean is also called expectation value;
\(\langle \hat \beta_\A \rangle = \beta_A\),
\(\langle \hat \beta_\B \rangle = \beta_B\), and therefore
\(\langle \widehat{\Delta \beta} \rangle = \Delta \beta = \beta_B - \beta_A\)
(cf.~Eq.~\ref{conest}).

For a true accuracy \(a\) the situation is not that simple, because we
do not have a generative model of the data \(y_t\) that is parametrized
by \(a\). But, consistent with the RFX model for accuracies and in
analogy to GLM parameters, we can define the true accuracy as the
expectation value of estimated accuracies,
\(a = \langle \hat a \rangle\), i.e.~the mean across an infinite number
of repetions of the experiment with the same subject. This true accuracy
\(a\) is a complex function of the true GLM parameters for the included
voxels and conditions, as well as the error variance and covariance, and
may also depend on the classification algorithm.

This definition is in line with the use of the term ‘true accuracy’ by
Brodersen et al. (2013). It goes beyond the definition of the true
accuracy of a particular trained classifier given by Pereira et al.
(2009) as the expectation of \(\hat a\) across test data sets (or
equivalently, the accuracy determined on an infinite amount of test
data). Like Brodersen and colleagues, we are not interested in
characterizing the performance of a classifier trained on specific
random training data, but in characterizing the subject as the source of
both training and test data.\footnote{Note that a classifier trained on
  a limited amount of training data will perform worse than optimally
  possible. This effect brings the accuracy closer to (but not below)
  chance level, i.e.~what we consider as the true accuracy is generally
  smaller than the ‘optimal true accuracy’:
  \(a_\mathrm{opt} \geq a \geq a_0\) (cf. Wyman et al., 1990).}

\subsection{Simulated distributions of contrasts and
accuracies}\label{simulation}

We here explain the basic outline of the simulations underlying
Sec.~\ref{do}; for full details of the implementation, see
App.~\ref{implementation}. The distributions arising in this simulation
are illustrated in Fig.~1, so that this appendix also serves as an
extended explanation of that figure.

Single-subject data are generated according to the first-level model of
Eq.~\ref{glm}. There are \(n = 25\) trials for each condition in each of
\(m = 6\) runs. At the second level, true activation differences
\(\Delta \beta_k\) for a sample of subjects are generated as independent
normally-distributed numbers with mean \(\Delta \mu\) and standard
deviation \(\sigma_2\) (Eq.~\ref{conpop}). A sample consists of
\(N = 17\) subjects. We used a relatively large number of trials and
runs in combination with trial-wise classification to obtain
fine-grained accuracy estimation distributions well-suited for graphical
display (red lines in Fig.~1e and f), and we chose a sample of 17
subjects to be able to closely approximate the standard significance
level of \(\alpha = 0.05\) for subject-wise classification (Fig.~2d).
The results remain qualitatively the same with other parameter choices
and for classification of run-wise parameter estimates. Since
multiplying all parameters by a constant factor changes only the scale
but not the structure of the data, we choose \(\sigma_1 = 1\). The
distributions of activation differences are illustrated in Figure~1a and
b, for the case \(\Delta \mu = 1\) and \(\sigma_2 = 4\) and for three
subjects with true contrasts \(\Delta \beta = -5, 1,\) and 8,
respectively.

Based on the generative model for single-subject data (Eq.~\ref{glm}),
the amount of information the data contain about the experimental
condition can be precisely calculated as a function of the true
activation difference \(\Delta \beta\) in that subject (Fig.~1c); it is
a symmetric function which reaches its minimum value of \(0\,\)bit for
\(\Delta \beta = 0\) and saturates towards \(1\,\)bit for large
\(|\Delta \beta|\). This demonstrates that in the calculation of
information the sign of the true activation difference is discarded, and
that there never is negative information.

The data are entered into a classification procedure for each subject
separately. Single-trial data from 5 of the 6 runs are used to train a
linear support vector machine (C-SVM with parameter \(C = 1\);
implementation by Chang and Lin, 2011, see
\url{http://www.csie.ntu.edu.tw/~cjlin/libsvm/}) which is applied to
trials from the left-out run, and this is repeated such that each run is
once used for testing (6-fold cross-validation). The proportion of
correctly classified trials gives the estimated classification accuracy
\(\hat a\). Fig.~1d shows the true accuracy \(a\) (calculated as the
mean of \(\hat a\) across many simulations) in a single subject as a
function of the true activation difference: \(a(\Delta \beta)\). The
shape of this function shows that classification accuracy is an
information-like measure: It is a symmetric function of \(\Delta \beta\)
with a minimum at \(\Delta \beta = 0\), where it reaches the ‘chance
level’ \(a_0 = 50\,\%\).

Actual estimation distributions of accuracies \(\hat a\) for the three
simulated subjects are shown in Fig.~1e (red lines). Note that in
contrast to the assumption of Eq.~\ref{accest}, these distributions are
not necessarily close to normal (or binomial; cf. Schreiber and
Krekelberg, 2013; Noirhomme et al., 2014; Jamalabadi et al., 2016) and
may not even be symmetric around the true value \(a\) (black bars); this
becomes particularly apparent for \(\Delta \beta = 1\) (Fig.~1e, S2).
However, the mean of the estimation distribution is by definition
identical to the true accuracy, which makes \(\hat a\) an unbiased
estimator of \(a\).

The function \(a(\Delta \beta)\) (Fig.~1d) associates each true
activation difference \(\Delta \beta\) with a corresponding true
accuracy \(a\). The population distribution of true activation
differences \(\Delta \beta_k\) (Fig.~1b, black line) therefore
determines the population distribution of true accuracies \(a_k\)
(Fig.~1f, black line). For the case illustrated here, this distribution
is far from normal (cf.~Eq.~\ref{accpop}); it is limited to
\(a \geq a_0\) but also has a spike at \(a = a_0\) due to the flat
minimum of \(a(\Delta \beta)\), while a weaker secondary maximum at
\(56\,\%\) is due to the nonlinear increase of that function. Because
\(\hat a\) is just an estimate of \(a\), the pronounced profile of this
population distribution \(a_k\) gives rise to a smoother distribution of
estimated accuracies across subjects \(\hat a_k\) (Fig.~1f, red line),
for which the normality assumption of Eq.~\ref{accpopest} might be
accepted as a rough approximation.

\subsection{Simulation implementation}\label{implementation}

We here fill in technical details of the numerical calculations and
simulations described in App.~\ref{simulation} and presented in
Figs.~1–3.

Single-subject data were simulated using the first-level GLM
(Eq.~\ref{glm}). For simplicity, each trial lasted for one time unit
(repetition time, ‘TR’), each time unit belonged to a trial of one of
the two conditions, and the hemodynamic response was assumed to be
instantaneous, i.e. \(x_{\A t}\) and \(x_{\B t}\) were sequences of 0s
and 1s. There were \(2 \cdot 25 \cdot 6 = 300\) time units in total
(conditions · trials per run · runs). A given activation difference
\(\Delta \beta\) was implemented by setting \(\beta_\A = 0\) and
\(\beta_\B = \Delta \beta\). The error \(e_t\) consisted of independent
(no temporal autocorrelation) normally-distributed pseudo-random numbers
with standard deviation \({\scriptstyle \sqrt\frac{m n}2} \sigma_1\), to
ensure that Eq.~\ref{conest} holds.

The mutual information (Cover and Thomas, 2012) between the data \(y\)
and the trial type \(T \in \{\A, \B\}\) encoded in the regressors shown
in \textbf{Fig.~1c} was calculated as \[
I(y, T) =  H(y) - H(y | T),
\] where the conditional entropy of \(y\) can be determined analytically
as \[
H(y | T) = \frac12 \log_2 (\pi e ~ m n),
\] but the marginal entropy \[
H(y) = \int - f_y(y) \log_2 f_y(y) ~ \ud y
\] was computed by numerical integration based on the marginal density
\[
f_y(y) = \frac12 \Dist N \left ( y ; 0, \frac{m n}2  \right )
+ \frac12 \Dist N \left ( y ; \Delta \beta, \frac{m n}2 \right ) .
\] Here \(\Dist N (x ; \mu, \sigma^2)\) denotes the density of the
normal distribution.

For the univariate classification results, data at the first level were
simulated for 100 values of the true contrast \(\Delta \beta\) from 0 to
86.6, at steps that were linearly increasing from 0.00883 to 1.74 to
achieve better coverage close to \(\Delta \beta = 0\). For each value,
400,000 time series \(y_t\) were randomly generated and the
classification accuracy \(\hat a\) was determined by cross-validation.
Histograms of the resulting estimation distribution of accuracy for
three different values of \(\Delta \beta\) were used for
\textbf{Fig.~1e}.

For each \(\Delta \beta\), the true accuracy \(a\) was estimated by the
mean, and the width of the estimation distribution \(\varsigma_1\) by
the standard deviation of generated values of \(\hat a\). To obtain
smooth functional relationships used for \textbf{Fig.~1d} and as a basis
for later calculations, both the mean and standard deviation were
modeled as functions of the contrast, \(a(\Delta \beta)\) and
\(\varsigma_1(\Delta \beta)\), using a cubic smoothing spline.

The population density of the true accuracy \(f_a(a)\) in
\textbf{Fig.~1f} (black line) was computed by transforming the normal
population density of the true contrast
\(f_{\Delta \beta}(\Delta \beta)\) (Eq.~\ref{conpop} and Fig.~1b, black
line) through the modeled function \(a(\Delta \beta)\) using the
standard rules of change of variables for densities, \[
f_a(a) = \left| \frac{d}{da} a_+^{-1}(a) \right |~ f_{\Delta \beta}(a_+^{-1}(a))
+ \left| \frac{d}{da} a_-^{-1}(a) \right |~ f_{\Delta \beta}(a_-^{-1}(a)),
\] where \(a_+^{-1}(a)\) and \(a_-^{-1}(a)\) denote the positive and
negative solution of \(a(\Delta \beta) = a\). The combined
population-and-estimation distribution of accuracies in Fig.~1f (red
line) was computed as an average of histograms of \(\hat a\) for
different values of \(\Delta \beta\), weighted by the population density
of \(\Delta \beta\).

The rejection probability of a second-level two-sided \(t\)-test on
contrast estimates as a function of \(\Delta \mu\) and \(\sigma_2\) in
\textbf{Fig.~2b} was computed as follows. Since the summary statistic
\(\widehat{\Delta \beta}_k\) is normally distributed with expectation
\(\Delta \mu\) and combined variance
\(\sigma^2_\mathrm{c} = \sigma^2_1 + \sigma^2_2\) (Eq.~\ref{conpopest}),
the \(t\)-statistic from \(N\) samples is noncentrally \(t\)-distributed
with \(N - 1\) degrees of freedom and noncentrality parameter
\(\Delta \mu \sqrt{N / \sigma^2_\mathrm{c}}\). The rejection probability
is the probability mass of this distribution exceeding the critical
value of the two-sided \(t\)-test with \(N - 1\) degrees of freedom in
either direction.

In a first approximation, the rejection probability of a second-level
one-sided \(t\)-test on accuracies vs \(a_0\) in \textbf{Fig.~2a} was
computed in the same way, by assuming that the distribution of the
summary statistic \(\hat a_k\) is exactly normal (Eq.~\ref{accpopest}
holds), with parameters \(\bar a\) and
\(\varsigma^2_\mathrm{c} = \varsigma^2_1 + \varsigma^2_2\). Note that we
do not assume this to result from a normal distribution of true
accuracies in the population (Eq.~\ref{accpop}), but we treat
Eq.~\ref{accpopest} solely as an approximate description of the actual
combined estimation and population variation (Fig.~1f, red line). Under
this approximation, the \(t\)-statistic from \(N\) samples is
noncentrally \(t\)-distributed with \(N - 1\) degrees of freedom and
noncentrality parameter
\((\bar a - a_0) \sqrt{N / \varsigma^2_\mathrm{c}}\). The distribution
moments of the summary statistic \(\hat a_k\) were numerically
calculated from the population distribution of \(\Delta \beta_k\)
(Eq.~\ref{conpop}) combined with the modeled parameters of the
estimation distribution of \(\hat a\) (Eq.~\ref{accest}),
\(a(\Delta \beta)\) and \(\varsigma_1(\Delta \beta)\), by evaluating \[
\begin{split}
\bar a &= \int a(\Delta \beta)
~ \Dist N (\Delta \beta ; \Delta \mu, \sigma^2_2) ~ \ud \Delta \beta
\quad \text{and} \\
\varsigma^2_c &= \int
\left ( \varsigma^2_1(\Delta \beta) + (a(\Delta \beta) - \bar a)^2 \right )
~ \Dist N (\Delta \beta ; \Delta \mu, \sigma^2_2) ~ \ud \Delta \beta.
\end{split}
\] The rejection probability is the probability mass of the resulting
distribution exceeding the critical value of the one-sided \(t\)-test
with \(N - 1\) degrees of freedom.

This approximation was finessed by comparing its results with those of a
simulation of \(t\)-tests at 26 values of \(\Delta \mu\) (0 to 0.5,
equidistant) combined with 101 values of \(\sigma_2\) (0 to 2,
equidistant). At each parameter setting, \(N = 17\) values of
\(\Delta \beta\) were sampled from the population distribution
(Eq.~\ref{conpop}), and for each a value of \(\hat a\) was drawn from
the corresponding estimation distribution. For this purpose, we did not
simulate time series data and classification again, but re-used the
result of the univariate classification simulation described above. For
each \(\Delta \beta\), the closest of the 100 values realized in that
simulation was determined, and \(\hat a\) was drawn randomly from the
pool of corresponding 400,000 simulation results. A one-sided \(t\)-test
vs \(a_0\) was applied to the \(N\) drawn accuracies and it was recorded
whether the null hypothesis could be rejected or not. This was repeated
5,000,000 times for each parameter setting. The resulting simulated
rejection probabilities were compared to the approximated rejection
probabilities and it was found that the approximation overestimates the
rejection probability for larger values (largest discrepancy 0.7270
instead of 0.7169) and underestimates it for smaller values (largest
discrepancy 0.1530 instead of 0.1604). The relation between simulated
and approximated rejection probabilities was modeled using a 4th-order
polynomial, which was then used to correct the approximation results for
Fig.~2a. This resulted in the smallest occurring approximated rejection
probability of 0.05 to be corrected to 0.055. The advantage of this
combination of approximation and simulation is that the semi-analytic
part provides a smooth function of the simulation parameters well-suited
for graphical display, which is guaranteed to be precise by calibrating
it using simulation.

The rejection probability of FFX analysis applied to contrast estimates
in \textbf{Fig.~2c} was computed as follows. FFX analysis also uses a
\(t\)-statistic, but it compares the mean estimated activation
difference \(\frac1N \sum_{k = 1}^N \widehat{\Delta \beta}_k\) not to an
estimate of the combined variance
\(\sigma^2_c = \sigma^2_1 + \sigma^2_2\), but of the estimation variance
\(\sigma^2_1\) only. \(\sigma^2_1\) is estimated from the first-level
GLM residuals with \(2 (m n - 1)\) degrees of freedom and the estimate
is pooled across \(N\) subjects, such that the resulting statistic is
\(t\)-distributed with \(2 (m n - 1) N\) degrees of freedom under the
FFX null hypothesis. However, under the RFX model including random
population variation, the variance of the statistic is actually larger
by a factor \(\sigma^2_c / \sigma^2_1\). For arbitrary \(\Delta \mu\),
the distribution of the FFX \(t\)-statistic is a scaled noncentral
\(t\)-distribution, with \(2 (m n - 1) N\) degrees of freedom,
noncentrality parameter \(\Delta \mu \sqrt{N / \sigma^2_\mathrm{c}}\),
and scaling factor \(\sigma_c / \sigma_1\). The rejection probability is
the probability mass of this distribution exceeding the critical value
of a two-sided \(t\)-test with \(2 (m n - 1) N\) degrees of freedom in
either direction.

The distribution of accuracies for classification across subjects
depends on the ratio \(\delta = 2 ~ \Delta \mu / \sigma_\mathrm{c}\)
(the subject-level Mahalanobis distance). For 100 values of \(\delta\)
from 0 to 10, at steps that were linearly increasing from 0.00102 to 0.2
to achieve better coverage close to 0, second-level univariate data
\(\hat \beta_{\A k}\) and \(\hat \beta_{\B k}\) were generated for
\(N = 17\) subjects. For 400,000 realizations, the resulting
classification accuracy was determined by leave-one-subject-out
cross-validation. The null hypothesis of no population difference
between conditions A and B is realized at \(\delta = 0\); from the
corresponding simulated null distribution the critical value was
determined to be an accuracy of \(67.6\,\%\). This gives the best
possible approximation of the 95th percentile of the discrete
distribution of accuracies with \(2 N + 1 = 35\) possible outcomes and
leads to a test at a significance level of \(\alpha = 0.051\). The
rejection probability for all simulated ratios was calculated as the
fraction of simulated accuracies that reach or exceed this value, and
the rejection probability was modeled as a function of \(\delta\) using
a cubic smoothing spline. The rejection probability of classification
across subjects in \textbf{Fig.~2d} was then computed by applying this
function to \[
\delta(\Delta \mu, \sigma_2) = 2 \frac{\Delta \mu}{\sqrt{\sigma^2_1 + \sigma^2_2}}
\] with \(\sigma^2_1 = 1\).

For the multivariate classification results, simulations were performed
in the same way as above, but with 10,000 simulated multi-voxel time
series for each value of the true contrast, with dimensions \(p = 2\)
and 10. In contrast to the univariate case, no semi-analytic
approximation was used, but the rejection probabilities in
\textbf{Fig.~3a} were directly estimated from 1,000,000 simulated
\(t\)-tests applied to accuracies resampled from this multivariate
simulation, for each of 100 equidistant values of \(\sigma_2\) from 0 to
2.

The rejection probabilities in \textbf{Fig.~3b} were directly estimated
from 5,000,000 simulated \(t\)-tests applied to univariate
classification accuracies. Again, the simulation of time series and
classification was not repeated, but the result of the univariate
classification simulation described above was re-used. \(N = 17\) values
of \(\Delta \beta\) were sampled from the alternative population model
of Eq.~\ref{simpleprop}, and for each a value of \(\hat a\) was drawn
from the corresponding estimation distribution. This was performed for
\(\Delta \beta^* = 1\), 2, and 5, and for 101 equidistant values of
\(\gamma\) from 0 to 1.

\section*{Note}\label{note}
\addcontentsline{toc}{section}{Note}

Though the spatially extended prevalence null hypothesis (PN) at
\(\gamma_0 = 0\) is equivalent to the spatially extended global null
hypothesis (GN), the stated significance criterion for PN
(cf.~Eq.~\ref{pcorprev}) \[
p^*_N(m | \gamma \leq \gamma_0) = p^*_N(m)
  + [1 - p^*_N(m)] p_N(m | \gamma \leq \gamma_0) \leq \alpha
\] at \(\gamma_0 = 0\), i.e. \[ \label{sigpn0}
p^*_N(m) + [1 - p^*_N(m)] p_N(m) \leq \alpha
\] is \emph{not} equivalent to the significance criterion for GN \[
p^*_N(m) \leq \alpha.
\] Rather, it is conservative due to the additional term
\([1 - p^*_N(m)] p_N(m)\).

This discrepancy carries over into the algorithm Step 5b, where the
criterion \(p_N(m) \leq \alpha^*\) (corresponding to Eq.~\ref{sigpn0})
for the prevalence lower bound (Eq.~\ref{confidence}) to be defined is
not equivalent to the significance criterion for GN.

However, if GN can be rejected (\(p^*_N(m) \leq \alpha\)), the effective
number of tests is moderately large (\textasciitilde{} 10), and a
standard significance level \(\alpha\) is used (e.g. \(0.05\)), it holds
\(p_N(m) \ll 1\), so that the difference between the criteria becomes
negligible.

Is summary, Eq.~\ref{pcorprev} is not exact but rather formulates (yet
another) upper bound, but the error introduced this way is small under
normal circumstances. In any case, the criterion is conservative and
therefore the test remains valid.

\section*{References}\label{references}
\addcontentsline{toc}{section}{References}

\hyperdef{}{refs}{\label{refs}}
\hyperdef{}{ref-AllefeldSearchlight}{\label{ref-AllefeldSearchlight}}
Allefeld, C., Haynes, J.-D., 2014. Searchlight-based multi-voxel pattern
analysis of fMRI by cross-validated MANOVA. Neuroimage 89, 345–357.

\hyperdef{}{ref-Bateson}{\label{ref-Bateson}}
Bateson, G., 1972. Steps to an ecology of mind. University of Chicago
Press, Chicago.

\hyperdef{}{ref-BrodersenVariational}{\label{ref-BrodersenVariational}}
Brodersen, K., Daunizeau, J., Mathys, C., Chumbley, J., Buhmann, J.,
Stephan, K., 2013. Variational Bayesian mixed-effects inference for
classification studies. Neuroimage 76, 345–361.

\hyperdef{}{ref-BrodersenBayesian}{\label{ref-BrodersenBayesian}}
Brodersen, K., Mathys, C., Chumbley, J., Daunizeau, J., Ong, C.,
Buhmann, J., Stephan, K., 2012. Bayesian mixed-effects inference on
classification performance in hierarchical data sets. The Journal of
Machine Learning Research 13, 3133–3176.

\hyperdef{}{ref-libsvm}{\label{ref-libsvm}}
Chang, C.-C., Lin, C.-J., 2011. LIBSVM: A library for support vector
machines. ACM Transactions on Intelligent Systems and Technology 2,
27:1–27:27.

\hyperdef{}{ref-CichyEncoding}{\label{ref-CichyEncoding}}
Cichy, R.M., Chen, Y., Haynes, J.-D., 2011. Encoding the identity and
location of objects in human LOC. Neuroimage 54, 2297–2307.

\hyperdef{}{ref-CoverElements}{\label{ref-CoverElements}}
Cover, T., Thomas, J., 2012. Elements of information theory. John Wiley
\& Sons, Hoboken.

\hyperdef{}{ref-CoxFunctional}{\label{ref-CoxFunctional}}
Cox, D., Savoy, R., 2003. Functional magnetic resonance imaging (fMRI)
’brain reading’: Detecting and classifying distributed patterns of fMRI
activity in human visual cortex. Neuroimage 19, 261–270.

\hyperdef{}{ref-DavisWhat}{\label{ref-DavisWhat}}
Davis, T., LaRocque, K., Mumford, J., Norman, K., Wagner, A., Poldrack,
R., 2014. What do differences between multi-voxel and univariate
analysis mean? How subject-, voxel-, and trial-level variance impact
fMRI analysis. Neuroimage 97, 271–283.

\hyperdef{}{ref-EfronBootstrap}{\label{ref-EfronBootstrap}}
Efron, B., Tibshirani, R., 1994. An introduction to the bootstrap.
Chapman \& Hall, London.

\hyperdef{}{ref-ErnstPermutation}{\label{ref-ErnstPermutation}}
Ernst, M., 2004. Permutation methods: A basis for exact inference.
Statistical Science 19, 676–685.

\hyperdef{}{ref-EtzelMVPA}{\label{ref-EtzelMVPA}}
Etzel, J., Braver, T., 2013. MVPA permutation schemes: Permutation
testing in the land of cross-validation, in: International Workshop on
Pattern Recognition in Neuroimaging. pp. 140–143.

\hyperdef{}{ref-EtzelSearchlight}{\label{ref-EtzelSearchlight}}
Etzel, J.A., Zacks, J.M., Braver, T.S., 2013. Searchlight analysis:
Promise, pitfalls, and potential. Neuroimage 78, 261–269.

\hyperdef{}{ref-FahrmeirRegression}{\label{ref-FahrmeirRegression}}
Fahrmeir, L., Kneib, T., Lang, S., Marx, B., 2013. Regression: Models,
methods and applications. Springer, Berlin.

\hyperdef{}{ref-FisherDesign}{\label{ref-FisherDesign}}
Fisher, R., 1935. The design of experiments. Oliver \& Boyd, Edinburgh.

\hyperdef{}{ref-FisherStatistical}{\label{ref-FisherStatistical}}
Fisher, R., 1925. Statistical methods for research workers. Oliver \&
Boyd, Edinburgh.

\hyperdef{}{ref-FristonMultisubject}{\label{ref-FristonMultisubject}}
Friston, K., Holmes, A., Price, C., Büchel, C., Worsley, K., 1999a.
Multisubject fMRI studies and conjunction analyses. Neuroimage 10,
385–396.

\hyperdef{}{ref-FristonSubjects}{\label{ref-FristonSubjects}}
Friston, K., Holmes, A., Worsley, K., 1999b. How many subjects
constitute a study? Neuroimage 10, 1–5.

\hyperdef{}{ref-FristonStatistical}{\label{ref-FristonStatistical}}
Friston, K., Holmes, A., Worsley, K., Poline, J.-P., Frith, C.,
Frackowiak, R., 1995. Statistical parametric maps in functional imaging:
A general linear approach. Human Brain Mapping 2, 189–210.

\hyperdef{}{ref-GaonkarAnalytic}{\label{ref-GaonkarAnalytic}}
Gaonkar, B., Davatzikos, C., 2013. Analytic estimation of statistical
significance maps for support vector machine based multi-variate image
analysis and classification. Neuroimage 78, 270–283.

\hyperdef{}{ref-GaonkarInterpreting}{\label{ref-GaonkarInterpreting}}
Gaonkar, B., Shinohara, R.T., Davatzikos, C., ADNI, 2015. Interpreting
support vector machine models for multivariate group wise analysis in
neuroimaging. Medical Image Analysis 24, 190–204.

\hyperdef{}{ref-GollandPermutation}{\label{ref-GollandPermutation}}
Golland, P., Fischl, B., 2003. Permutation tests for classification:
Towards statistical significance in image-based studies, in: Information
Processing in Medical Imaging. Springer, Berlin, pp. 330–341.

\hyperdef{}{ref-GoergenSAA}{\label{ref-GoergenSAA}}
Görgen, K., Hebart, M., Allefeld, C., Haynes, J.-D., 2014. Detecting,
avoiding \& eliminating confounds in MVPA / decoding studies. Poster
presented at the 20th annual meeting of the Organization for Human Brain
Mapping (OHBM). available at
http://dx.doi.org/10.7490/f1000research.1111808.1.

\hyperdef{}{ref-HaufeInterpretation}{\label{ref-HaufeInterpretation}}
Haufe, S., Meinecke, F., Görgen, K., Dähne, S., Haynes, J.-D.,
Blankertz, B., Bießmann, F., 2014. On the interpretation of weight
vectors of linear models in multivariate neuroimaging. Neuroimage 87,
96–110.

\hyperdef{}{ref-HaxbyEarly}{\label{ref-HaxbyEarly}}
Haxby, J., 2012. Multivariate pattern analysis of fMRI: The early
beginnings. Neuroimage 62, 852–855.

\hyperdef{}{ref-HaxbyDistributed}{\label{ref-HaxbyDistributed}}
Haxby, J., Gobbini, M., Furey, M., Ishai, A., Schouten, J., Pietrini,
P., 2001. Distributed and overlapping representations of faces and
objects in ventral temporal cortex. Science 293, 2425–2430.

\hyperdef{}{ref-HaxbyCommon}{\label{ref-HaxbyCommon}}
Haxby, J., Guntupalli, J., Connolly, A., Halchenko, Y., Conroy, B.,
Gobbini, M., Hanke, M., Ramadge, P., 2011. A common, high-dimensional
model of the representational space in human ventral temporal cortex.
Neuron 72, 404–416.

\hyperdef{}{ref-HaynesDecoding}{\label{ref-HaynesDecoding}}
Haynes, J.-D., Rees, G., 2006. Decoding mental states from brain
activity in humans. Nature Reviews Neuroscience 7, 523–534.

\hyperdef{}{ref-HaynesConsciousness}{\label{ref-HaynesConsciousness}}
Haynes, J.-D., Rees, G., 2005a. Predicting the stream of consciousness
from activity in human visual cortex. Current Biology 15, 1301–1307.

\hyperdef{}{ref-HaynesOrientation}{\label{ref-HaynesOrientation}}
Haynes, J.-D., Rees, G., 2005b. Predicting the orientation of invisible
stimuli from activity in human primary visual cortex. Nature
Neuroscience 8, 686–691.

\hyperdef{}{ref-HaynesReading}{\label{ref-HaynesReading}}
Haynes, J.-D., Sakai, K., Rees, G., Gilbert, S., Frith, C., Passingham,
R., 2007. Reading hidden intentions in the human brain. Current Biology
17, 323–328.

\hyperdef{}{ref-HebartDecoding}{\label{ref-HebartDecoding}}
Hebart, M., Görgen, K., Haynes, J.-D., 2015. The Decoding Toolbox (TDT):
A versatile software package for multivariate analyses of functional
imaging data. Frontiers in Neuroinformatics 8, 88.

\hyperdef{}{ref-HolmesGeneralisability}{\label{ref-HolmesGeneralisability}}
Holmes, A., Friston, K., 1998. Generalisability, random effects \&
population inference. Neuroimage 7, S754.

\hyperdef{}{ref-Hoyos-IdroboImproving}{\label{ref-Hoyos-IdroboImproving}}
Hoyos-Idrobo, A., Schwartz, Y., Varoquaux, G., Thirion, B., 2015.
Improving sparse recovery on structured images with bagged clustering,
in: International Workshop on Pattern Recognition in Neuroimaging. pp.
73–76.

\hyperdef{}{ref-HuberRobust}{\label{ref-HuberRobust}}
Huber, P., Ronchetti, E., 2009. Robust statistics, 2nd ed. John Wiley \&
Sons, Hoboken.

\hyperdef{}{ref-JamalabadiClassification}{\label{ref-JamalabadiClassification}}
Jamalabadi, H., Alizadeh, S., Schönauer, M., Leibold, C., Gais, S.,
2016. Classification based hypothesis testing in neuroscience:
Below-chance level classification rates and overlooked statistical
properties of linear parametric classifiers. Human Brain Mapping 37,
1842–1855.

\hyperdef{}{ref-KowalczykClassification}{\label{ref-KowalczykClassification}}
Kowalczyk, A., 2007. Classification of anti-learnable biological and
synthetic data, in: Knowledge Discovery in Databases: PKDD 2007.
Springer, Berlin, pp. 176–187.

\hyperdef{}{ref-KriegeskorteAnalyzing}{\label{ref-KriegeskorteAnalyzing}}
Kriegeskorte, N., Bandettini, P., 2007. Analyzing for information, not
activation, to exploit high-resolution fMRI. Neuroimage 38, 649–662.

\hyperdef{}{ref-KriegeskorteInformation}{\label{ref-KriegeskorteInformation}}
Kriegeskorte, N., Goebel, R., Bandettini, P., 2006. Information-based
functional brain mapping. Proceedings of the National Academy of
Sciences of the United States of America 103, 3863–3868.

\hyperdef{}{ref-LazarCombining}{\label{ref-LazarCombining}}
Lazar, N., Luna, B., Sweeney, J., Eddy, W., 2002. Combining brains: A
survey of methods for statistical pooling of information. Neuroimage 16,
538–550.

\hyperdef{}{ref-LehmannTesting}{\label{ref-LehmannTesting}}
Lehmann, E., Romano, J., 2005. Testing statistical hypotheses, 3rd ed.
Springer, Berlin.

\hyperdef{}{ref-Mourao-MirandaClassifying}{\label{ref-Mourao-MirandaClassifying}}
Mourao-Miranda, J., Bokde, A.L.W., Born, C., Hampel, H., Stetter, M.,
2005. Classifying brain states and determining the discriminating
activation patterns: Support Vector Machine on functional MRI data.
Neuroimage 28, 980–995.

\hyperdef{}{ref-MudholkarLogit}{\label{ref-MudholkarLogit}}
Mudholkar, G., George, E., 1979. The logit method for combining
probabilities, in: Symposium on Optimizing Methods in Statistics.
Academic Press, New York, pp. 345–366.

\hyperdef{}{ref-NicholsValid}{\label{ref-NicholsValid}}
Nichols, T., Brett, M., Andersson, J., Wager, T., Poline, J.-B., 2005.
Valid conjunction inference with the minimum statistic. Neuroimage 25,
653–660.

\hyperdef{}{ref-NicholsNonparametric}{\label{ref-NicholsNonparametric}}
Nichols, T., Holmes, A., 2001. Nonparametric permutation tests for
functional neuroimaging: A primer with examples. Human Brain Mapping 15,
1–25.

\hyperdef{}{ref-NiliToolbox}{\label{ref-NiliToolbox}}
Nili, H., Wingfield, C., Walther, A., Su, L., Marslen-Wilson, W.,
Kriegeskorte, N., 2014. A toolbox for representational similarity
analysis. PLoS Computational Biology 10, e1003553.

\hyperdef{}{ref-NoirhommeBiased}{\label{ref-NoirhommeBiased}}
Noirhomme, Q., Lesenfants, D., Gomez, F., Soddu, A., Schrouff, J.,
Garraux, G., Luxen, A., Phillips, C., Laureys, S., 2014. Biased binomial
assessment of cross-validated estimation of classification accuracies
illustrated in diagnosis predictions. Neuroimage: Clinical 4, 687–694.

\hyperdef{}{ref-NormanBeyond}{\label{ref-NormanBeyond}}
Norman, K., Polyn, S., Detre, G., Haxby, J., 2006. Beyond mind-reading:
Multi-voxel pattern analysis of fMRI data. Trends in Cognitive Sciences
10, 424–430.

\hyperdef{}{ref-OlivettiBayesian}{\label{ref-OlivettiBayesian}}
Olivetti, E., Veeramachaneni, S., Nowakowska, E., 2012. Bayesian
hypothesis testing for pattern discrimination in brain decoding. Pattern
Recognition 45, 2075–2084.

\hyperdef{}{ref-PennyRandom}{\label{ref-PennyRandom}}
Penny, W., Holmes, A., 2007. Random effects analysis, in: Friston, K.,
others (Eds.), Statistical Parametric Mapping. Academic Press, London,
pp. 156–165.

\hyperdef{}{ref-PennyRandomOld}{\label{ref-PennyRandomOld}}
Penny, W., Holmes, A., 2004. Random-effects analysis, in: Frackowiak,
R., others (Eds.), Human Brain Function. Academic Press, London, pp.
843–850.

\hyperdef{}{ref-PereiraInformation}{\label{ref-PereiraInformation}}
Pereira, F., Botvinick, M., 2011. Information mapping with pattern
classifiers: A comparative study. Neuroimage 56, 476–496.

\hyperdef{}{ref-PereiraMachine}{\label{ref-PereiraMachine}}
Pereira, F., Mitchell, T., Botvinick, M., 2009. Machine learning
classifiers and fMRI: A tutorial overview. Neuroimage 45, S199–S209.

\hyperdef{}{ref-PriceCognitive}{\label{ref-PriceCognitive}}
Price, C., Friston, K., 1997. Cognitive conjunction: A new approach to
brain activation experiments. Neuroimage 5, 261–270.

\hyperdef{}{ref-RaschRobust}{\label{ref-RaschRobust}}
Rasch, D., Guiard, V., 2004. The robustness of parametric statistical
methods. Psychology Science 46, 175–208.

\hyperdef{}{ref-RosenblattRevisiting}{\label{ref-RosenblattRevisiting}}
Rosenblatt, J., Vink, M., Benjamini, Y., 2014. Revisiting multi-subject
random effects in fMRI: Advocating prevalence estimation. Neuroimage 84,
113–121.

\hyperdef{}{ref-RouderDetecting}{\label{ref-RouderDetecting}}
Rouder, J., Morey, R., Speckman, P., Pratte, M., 2007. Detecting chance:
A solution to the null sensitivity problem in subliminal priming.
Psychonomic Bulletin \& Review 14, 597–605.

\hyperdef{}{ref-SabuncuUniversal}{\label{ref-SabuncuUniversal}}
Sabuncu, M.R., 2014. A universal and efficient method to compute maps
from image-based prediction models, in: Medical Image Computing and
Computer-Assisted Intervention. Springer, Berlin, pp. 353–360.

\hyperdef{}{ref-SabuncuRelevance}{\label{ref-SabuncuRelevance}}
Sabuncu, M.R., Van Leemput, K., 2012. The relevance voxel machine
(RVoxM): A self-tuning Bayesian model for informative image-based
prediction. IEEE Transactions on Medical Imaging 31, 2290–2306.

\hyperdef{}{ref-SchreiberStatistical}{\label{ref-SchreiberStatistical}}
Schreiber, K., Krekelberg, B., 2013. The statistical analysis of
multi-voxel patterns in functional imaging. PLOS one 8, e69328.

\hyperdef{}{ref-SearleVariance}{\label{ref-SearleVariance}}
Searle, S., Casella, G., McCulloch, C., 1992. Variance components. John
Wiley \& Sons, Hoboken.

\hyperdef{}{ref-SochGLM}{\label{ref-SochGLM}}
Soch, J., Haynes, J.-D., Allefeld, C., 2016. How to avoid mismodelling
in GLM-based fMRI data analysis: Cross-validated Bayesian model
selection. Neuroimage.
doi:\href{http://doi.org/10.1016/j.neuroimage.2016.07.047}{10.1016/j.neuroimage.2016.07.047}

\hyperdef{}{ref-SpiridonDistributed}{\label{ref-SpiridonDistributed}}
Spiridon, M., Kanwisher, N., 2002. How distributed is visual category
information in human occipito-temporal cortex? An fMRI study. Neuron 35,
1157–1165.

\hyperdef{}{ref-StelzerStatistical}{\label{ref-StelzerStatistical}}
Stelzer, J., Chen, Y., Turner, R., 2013. Statistical inference and
multiple testing correction in classification-based multi-voxel pattern
analysis (MVPA): Random permutations and cluster size control.
Neuroimage 65, 69–82.

\hyperdef{}{ref-StephanBayesian}{\label{ref-StephanBayesian}}
Stephan, K., Penny, W., Daunizeau, J., Moran, R., Friston, K., 2009.
Bayesian model selection for group studies. Neuroimage 46, 1004–1017.

\hyperdef{}{ref-StoufferAmerican}{\label{ref-StoufferAmerican}}
Stouffer, S., Lumsdaine, A., Lumsdaine, M., Williams Jr., R., Smith, M.,
Janis, I., Star, S., Cottrell Jr., L., 1949. The American soldier:
Combat and its aftermath. Princeton University Press, Princeton.

\hyperdef{}{ref-ThirionShortcomings}{\label{ref-ThirionShortcomings}}
Thirion, B., Flandin, G., Pinel, P., Roche, A., Ciuciu, P., Poline,
J.-B., 2006. Dealing with the shortcomings of spatial normalization:
Multi-subject parcellation of fMRI datasets. Human Brain Mapping 27,
678–693.

\hyperdef{}{ref-TippettMethods}{\label{ref-TippettMethods}}
Tippett, L., 1931. The methods of statistics. Williams \& Norgate,
London.

\hyperdef{}{ref-ToddConfounds}{\label{ref-ToddConfounds}}
Todd, M., Nystrom, L., Cohen, J., 2013. Confounds in multivariate
pattern analysis: Theory and rule representation case study. Neuroimage
77, 157–165.

\hyperdef{}{ref-WangSupport}{\label{ref-WangSupport}}
Wang, Z., Childress, A.R., Wang, J., Detre, J.A., 2007. Support vector
machine learning-based fMRI data group analysis. Neuroimage 36,
1139–1151.

\hyperdef{}{ref-WorsleyTest}{\label{ref-WorsleyTest}}
Worsley, K., Friston, K., 2000. A test for a conjunction. Statistics \&
Probability Letters 47, 135–140.

\hyperdef{}{ref-WymanAsymptotic}{\label{ref-WymanAsymptotic}}
Wyman, F., Young, D., Turner, D., 1990. A comparison of asymptotic error
rate expansions for the sample linear discriminant function. Pattern
Recognition 23, 775–783.

\end{document}